# Photosynthetic energy transfer: missing in action (detected spectroscopy)?

Ariba Javed[1,2†], Julian Lüttig[1†], Kateřina Charvátová[3], Stephanie E. Sanders[1], Rhiannon Willow[1], Muyi Zhang[1], Alastair T. Gardiner[4], Pavel Malý[3*] and Jennifer P. Ogilvie[1*]

[1]Department of Physics, University of Michigan, 450 Church St, Ann Arbor MI 48109, USA

[2]Department of Materials Science and Engineering, University of Michigan, 2300 Hayward St., Ann Arbor, Michigan 48109-2136, USA

[3]Faculty of Mathematics and Physics, Charles University, Ke Karlovu 5, 121 16 Praha 2, Czech Republic

[4]Center Algatech, Institute of Microbiology, Czech Academy of Sciences, 37901 Třeboň, Czech Republic

† Equal author contributions

*Corresponding authors: Jennifer P. Ogilvie (experiment) and Pavel Malý (theory)

**Email:** jogilvie@umich.edu, pavel.maly@matfyz.cuni.cz

## Abstract

In recent years, action-detected ultrafast spectroscopies have gained popularity. These approaches offer some advantages over their coherently-detected counterparts, enabling spatially-resolved and *operando* measurements with high sensitivity. However, there are also fundamental limitations connected to the different process of signal generation in action-detected experiments. Specifically, state mixing by nonlinear interactions during signal emission leads to a large static background which can obscure the excited-state dynamics. This could severely limit the applicability of action-detected spectroscopy to study energy transfer in larger systems. Here we perform fluorescence-detected two-dimensional electronic spectroscopy (F-2DES) of the light-harvesting II (LH2) complex from purple bacteria. We demonstrate that the B800-B850 energy transfer process in LH2 is barely discernable in F-2DES, representing a ~6.2% rise of the lower cross-peak intensity. This is in stark contrast to measurements using coherently-detected 2DES where the lower cross-peak reveals energy transfer with 100% contrast. We explain the weak presence of excited-state dynamics using a disordered excitonic model with realistic experimental conditions. We further derive a general formula for the presence of excited-state signals in multi-chromophoric aggregates, dependent on the aggregate geometry and size, and the interplay of excitonic coupling and disorder. We find that, dependent on the excitonic state structure, the excited state dynamics in F-2DES can be visible even in large aggregates. Our work shows that the signatures of energy transfer in F-2DES can be used to directly infer the excitonic structure in multichromophoric systems.

**Significance Statement**

Two-dimensional electronic spectroscopy (2DES) has proven to be a powerful approach for elucidating electronic structure and ultrafast dynamics. Recently, action-detected variants of 2DES employing fluorescence, photocurrent, photoions and photoelectrons have been developed, enabling spatially-resolved and *operando* measurements. The excited-state populations measured in action-detected experiments live orders of magnitude longer than the coherences detected in conventional coherently-detected 2DES. Using the light-harvesting II complex (LH2) as a model system we demonstrate that energy transfer is barely discernable in F-2DES as a result of dominant excited-state-lifetime-limited signals that arise from the multichromophoric nature of LH2. We show that weak signatures of energy transfer in LH2 and other aggregates are directly linked to excitonic state structure, making F-2DES a unique probe of excitonic characteristics.

**Main Text**

**Introduction**

Two-dimensional electronic spectroscopy (2DES) has emerged as a powerful tool for studying electronic structure and dynamics in systems ranging from photosynthetic complexes to liquids and solid-state materials (1, 2). Both high temporal and spectral resolution can be achieved using Fourier transform 2DES, providing a significant advantage over lower dimensional approaches such as pump-probe spectroscopy. In addition, by expressing the signal as a 2D map correlating the excitation and detection frequencies, 2DES effectively decongests the spectral information available from conventional linear spectroscopies, enabling a better understanding of a wide variety of systems (3). The most common implementation of 2DES utilizes a sequence of three time-delayed pulses that interact with the sample, generating a macroscopic third-order polarization that radiates a coherent signal field. Phase-matching is typically used to detect the coherent signal in a specific direction, requiring a sample volume greater than $\sim\lambda^3$ (where λ is the excitation wavelength) to attain the coherent build-up of a macroscopic signal (4). This limits the use of coherent 2DES to studies of extended samples containing large numbers of molecules, making it ill-suited for single molecule studies and for combining with microscopy.

In recent years, motivated by the desire to make spatially-resolved measurements and to correlate optical excitation with a wide range of observables, action-detected variants of 2DES and pump-probe spectroscopy have been developed that employ a fully collinear geometry, relying on phase-cycling or phase-modulation to extract the signals of interest. These approaches overcome some of the limitations of coherent 2DES and have employed fluorescence (5-7), photocurrent (8-11), photoelectron (12-14), or photoion (15, 16) emission as experimental observables. Fluorescence-detection, in general, has enabled spatially-resolved measurements, while photocurrent, photoion and photoelectron detection have enabled *operando* device studies (17). Numerous interesting systems including photosynthetic complexes (6, 7), atomic vapors (5, 13), molecular dimers (18, 19), dyads (20), semiconductor nanostructures (8, 10, 21, 22), as well as photovoltaic devices (11, 17, 23) have been studied using action-detected 2DES techniques. However, there are significant fundamental differences between action-detected and coherent methods, and the interpretation of the spectra obtained from action-detected 2DES still remains an active research topic (24-30). Studies comparing the spectra of coherently and action-detected measurements have been reported (19) (31, 32), but to date there have been relatively few studies of excited state dynamics with action-detected methods. In this work, we focus on fluorescence-detected 2DES (F-2DES) to highlight the similarities and differences between coherent and action-detected 2DES techniques for studying multichromophoric systems.

Compared to coherent 2DES (C-2DES), F-2DES adds a fourth pulse, projecting the third-order polarization onto an excited-state population of fourth order in the electric field. The resulting observable is an incoherent signal proportional to this excited-state population (5). The difference between C-2DES and F-2DES arises due to the fourth pulse which leads to additional Liouville pathways (18, 26, 30). The double-sided Feynman diagrams corresponding to these pathways are shown in *SI Appendix* Fig. S1, S2. In C-2DES, three contributions make up the spectra – the ground state bleach (GSB), stimulated emission (SE), and excited-state absorption (ESA), with the ESA pathway contributing with an opposite sign relative to the GSB and SE pathways. In F-2DES, in addition to GSB and SE pathways, two distinct and oppositely-signed ESA pathways exist, arising from emission from either the singly or doubly excited states. Depending on the relative quantum yields of the two excited states, the ESA pathways may partially or fully cancel, dramatically changing the appearance of F-2DES relative to C-2DES (18, 30).

The measured excited-state populations in molecular systems typically have lifetimes on the scale of nanoseconds, orders of magnitude longer-lived than electronic coherences. This long-lasting signal generation provides additional opportunities for excited-state populations to interact and become correlated over longer timescales leading to partial or even full cancellation of the ESA pathways. Second-order decay processes such as exciton-exciton annihilation (EEA), Auger recombination, photocarrier scattering, or nonlinearities in the detection can occur, contaminating the nonlinear signal and obscuring spectral dynamics (25, 28, 33, 34). Using F-2DES, strong cross peaks at early waiting times have been reported in multichromophoric systems (6, 7), in contrast to the observations in C-2DES studies where weak or no cross peaks were present (35). The appearance of cross peaks in F-2DES at early waiting times has been attributed to EEA (26, 30, 36) rather than to excitonic coupling (37) as in C-2DES. F-2DES has revealed clear excited-state dynamics in small systems such as molecular dyads (18-20) and Rb atoms (16). However, excited-state dynamics in extended multichromophoric systems have not been investigated, with the exception of weak kinetic signatures reported in photocurrent-detected 2DES studies of an organic photovoltaic device (9). Recent theoretical work suggests that action-detected 2DES is poorly suited for this task (29).

Here we study the light-harvesting II (LH2) complex from the purple bacterium *Rhodoblastus acidophilus (R. acidophilus)* using C-2DES and F-2DES to compare the ability of both techniques to resolve excited-state dynamics in multichromophoric systems. The well-characterized structure and the extensive previous ultrafast studies (38-44) of LH2 make it an excellent model system for this purpose. It is comprised of two bacteriochlorophyll *a* (*BChl a*) rings, B800 and B850. The B800 ring consists of 9 monomeric *BChl a*, which are weakly coupled to each other giving rise to an absorption peak at ~800 nm. The B850 ring consists of 18 strongly coupled *BChl a* arranged in dimeric units, producing an absorption peak at ~850 nm. *SI Appendix* Fig. S5 shows the structure and orientation of the B800 and B850 rings in LH2 from *R. acidophilus* (45). Many previous experimental studies have used C-2DES to probe the excited state dynamics within LH2, and have reported timescales ranging from ~700 fs to 1 ps for the B800 → B850 energy transfer (38-41). Previous F-2DES studies of LH2 studies showed spectra at $t_2$ = 0 but did not report waiting-time-dependent F-2DES measurements (6, 7).

To gain insight into the F-2DES measurements of LH2 we formulate an excitonic model-based description of the F-2DES response. We derive a general dependence of the response on the excitonic structure, allowing us to quantify the effects of varying oscillator strength, excited state population and excitonic delocalization on the prominence of the excited-state dynamics in the F-2DES spectra. The model successfully reproduces the measured LH2 F-2DES spectra and the weak B800-B850 energy-transfer signatures. We also examine other multichromophoric geometries, finding that the prominence of excited-state dynamics in F-2DES measurements is dependent upon the excitonic state structure and the combined effect of excitonic delocalization and finite-bandwidth laser spectra.

**Results**

The absorptive F-2DES spectra of LH2 are shown in Fig. 1 (top row) at waiting times $t_2$= 0 fs, 500 fs and 2 ps. The $t_2$ = 0 fs spectra show strong B800 and B850 diagonal peaks and prominent cross-peaks, consistent with previous work (6, 7). By 500 fs, the relative amplitudes have changed, with the upper diagonal peak (UDP) from B800 losing intensity relative to the lower diagonal peak (LDP) from B850, and the ratio of cross-peak/diagonal peak amplitude increasing. Overall, the F-2DES spectra show little evolution in peak amplitudes or shapes as a function of waiting time, with the lower cross-peak (LCP), which corresponds to excitation of B800 and detection of B850, showing the largest evolution in peak shape. The simulated F-2DES spectra are shown in Fig. 1 (middle row), displaying good agreement with the experimental measurements. In contrast to the F-2DES spectra, Fig. 1 (bottom row) shows the C-2DES spectra of LH2 at the same waiting times, revealing dramatic spectral evolution with increasing waiting time.

Fig. 2 shows the waiting-time dependence of the F-2DES diagonal and cross-peak amplitudes. The colored squares overlaid on the $t_2$ = 0 fs of the experimental and theoretical F-2DES spectra in Fig. 1 indicate the region over which the signal was integrated to obtain the kinetic traces shown in Fig. 2. All four peak amplitudes show a dramatic drop within the pulse overlap region, followed by very small amplitude changes within the first 2 ps. As we confirm by calculations, the rapid initial signal drop reflects coherence dephasing that we discuss further in the *SI Appendix*. The excitation dynamics are visible in the small changes that are more apparent in the panels below each kinetic trace, where the vertical scale is expanded. An exponential fit of the decay of the UDP amplitude reveals a time constant of 1480 ± 375 fs, and corresponds to a change of the signal amplitude of 4.4%. The LDP shows a rapid decay within the first ~200 fs followed by a roughly constant signal level. The upper cross-peak (UCP) shows little evolution during the first 2 ps aside from the initial amplitude drop. The LCP exhibits a small increase, which is fit to an exponential rise yielding a time constant of 1345 ± 575 fs, and corresponds to a signal rise of 6.2%. The corresponding waiting-time dependence of the C-2DES diagonal and cross-peak amplitudes are shown in *SI Appendix* Fig. S10. In contrast to F-2DES, the peak amplitude changes in C-2DES are much more apparent. In particular, the LCP, which is negligible at early waiting times, shows a clear exponential rise with a time constant of 825 ± 10 fs, amounting to ~100% of the cross-peak amplitude. Further details about the data fitting are provided in the *SI Appendix*.

Bolzonello et al. recently presented theoretical studies of action-detected 2DES of molecular assemblies (29). They find that when the output signal intensity is not proportional to the number of excitations generated in the system, *e.g.* because of EEA, the visibility of excited-state dynamics is reduced as the number of chromophores increases. When the EEA is efficient, as is the typical case of molecular aggregates such as LH2, the intensity of excited state dynamics in the signal is given by the ratio of the SE to GSB contribution. Based on a combinatoric argument, the SE/GSB ratio scales as 1/*N* for an aggregate of *N* identical molecules. Here we test the applicability of their combinatoric approach to LH2. In Fig. 3 we present double-sided Feynman diagrams that depict the signals that contribute to the LCP and UDP of the F-2DES spectrum of LH2; the features sensitive to the B800-B850 energy transfer. Only the SE and GSB pathways are present since the two oppositely-signed ESA contributions (shown in *SI Appendix* Fig. S1, S2) cancel (20). For this simple combinatoric argument we consider the system to be weakly coupled, such that the site basis is appropriate, with the number of pigments on the B800 and B850 rings given by $N_{800}$ = 9 and $N_{850}$ = 18 respectively. We later discuss the implications of excitonic coupling.

The LCP consists of diagrams in which the first pulse pair interacts with the $i^{th}$ pigment on the B800 ring, while the second pulse pair interacts with the $j^{th}$ pigment on B850 ring as shown in Fig. 3a). The SE pathways clearly exhibit interband energy transfer from B800 to B850 during the waiting time $t_2$, resulting in a rise in the LCP amplitude. In addition to the SE pathways, an even larger number of GSB pathways exist that contribute to the LCP. The number of SE pathways is $N_{B800}$ = 9. In contrast, the number of GSB pathways is $N_{B800} * N_{B850}$ = 9*18 = 162. If all pathways were equally weighted, the fraction of signal showing kinetics would be given by SE/GSB = 9/162 = 1/18 = 5.6% of the signal. This appears to be in quite good agreement with the data shown in Fig. 2d) in which the exponential rise due to B800-B850 energy transfer constitutes ~6.2% of the signal.

A similar argument can be made for the UDP, which has contributions from the GSB and SE pathways shown in Fig. 3b). In these pathways, the initial pulse pair interacts with the $i^{th}$ B800 pigment, while the second pulse pair interacts with the $k^{th}$ B800 pigment. The spectroscopic signatures of interband energy transfer, i.e., a rise of the LCP and a decay of the UDP, are connected to all the SE pathways since only these pathways include an excited state population during the waiting time $t_2$. The, intraband energy transfer can occur during waiting time $t_2$ as well. However, stimulated emission must occur from whichever pigment is in the excited state at the time of the third light-matter interaction (the $k^{th}$ pigment), making the total number of SE pathways equal to $N_{B800}$ = 9. In contrast, there are $N_{B800}* N_{B800}$ = 81 GSB pathways including the correlation of all B800 transitions. Assuming equal contributions of all pathways, for the UDP, the ratio of SE/GSB = 9/81 = 1/9 = 11%. This does not agree with the data, which show a considerably lower prominence of excited-state kinetics of ~4.4% derived from the exponential fit to the UDP in Fig. 2b).

From the pathway analysis above, it is clear that the excited-state dynamics is reflected in the SE contribution, visible as the UDP decay and LCP rise with the energy transfer between the rings. This transfer makes it possible to distinguish the spectrally identical SE and GSB contributions. Since only the number of pathways figures into the combinatoric argument, the disagreement with experiment suggests that the assumption of equal weight of all pathways is not valid. This implies different transition strengths in aggregates composed of the same molecules (*BChl a* in case of LH2), indicative of excitonic effects such as delocalization.

To understand the experimental F-2DES data and the effect of excitonic structure, we performed simulations to analyze the nonlinear response using Liouville-space pathways and account for the excitonic nature of LH2. As before, we assume, due to EEA, that the spectra contain only two types of contribution: GSB and SE, of which only the latter reports on the excited-state dynamics. We employ a standard disordered excitonic model for LH2 (46-48) as described in the Methods section and in detail in the *SI Appendix*. Diagonalization of the system Hamiltonian yields the excitonic state structure shown in Fig. 3c). In the B800 ring, the excitons are mostly localized, with similar oscillator strengths. In contrast, in the B850 ring a few bright states dominate the absorption band, while the rest have small oscillator strength and higher energy. The linear absorption (Fig. 3c) and F-2DES spectra (Fig. 1 Middle row) were calculated by summing all the appropriate response pathways (see *SI Appendix*). To better represent the experiment, the calculated F-2DES spectra of LH2 were additionally multiplied by the laser spectrum along both frequency axes and convolved with the Fourier transform of the time-domain filter function used in the experimental data processing. As can be seen from Fig. 1, the agreement of the theory and experiment is excellent. Upon applying the same analysis that we used for the experimental data, we extract the apparent SE/GSB ratios of the B800 and B850 manifolds directly from the kinetics of the integrated UDP and LCP as shown in Fig 2b) and d) respectively. We obtain contrasts of SE/GSB = 4.1% and SE/GSB = 4.8% from the UDP and LCP, in good agreement with the experimental values of 4.4% and 6.2% respectively. This agreement demonstrates that the combination of excitonic effects and the influence of finite laser spectra on the measured line-shapes can explain the observed presence of excited-state dynamics in F-2DES measurements.

To disentangle the effects of excitonic structure and the finite laser spectrum we calculated the F-2DES spectra with spectrally flat pulses. In Figure S14 we show that the finite laser spectrum effectively pulls the diagonal and cross-peaks together, making their individual kinetics more difficult to separate. Fig. 3d) compares the excited-state kinetics of the UDP decay and LCP rise calculated with the finite laser bandwidth (dashed lines), flat laser spectrum (solid lines), and the 1/*N* limit that assumes identical sites and perfect distinction of the SE and GSB. Clearly, the finite bandwidth increases the peak overlap, decreasing the apparent SE/GSB ratio and thus the excited-state dynamics visibility. Crucially, as seen from the LCP rise for the flat spectra, the excitonic effects can enhance the excited state dynamics far beyond what is expected from the number of pigments. To gain insight into this effect, we consider a generic aggregate of *N* coupled two-level systems. The response pathways can be separated into those with excitonic population and inter-exciton coherence in $t_2$. The inter-excitonic coherence dephases rapidly and leads to the initial signal drop which is not relevant for the excitation transfer. We thus consider only the population pathways. The ratio between the spectrally-integrated SE and GSB in the absorptive F-2DES spectra can be expressed as

$$\frac{\text{SE}}{\text{GSB}} = \frac{\sum_{i,j=1}^{N} \langle \mu_i \mu_i \mu_j \mu_j \rangle U_{ji}(t_2)}{\sum_{i,j=1}^{N} \langle \mu_i \mu_i \mu_j \mu_j \rangle}. \quad (1)$$

Here, $\mu_i$ is the transition dipole moment of the excitonic state $i$, and $U_{ji}(t_2)$ is the population propagator in the waiting time $t_2$, i.e., the conditional probability that, provided the excitation started in state $i$, it will be in state $j$ after time $t_2$ has passed. The orientational averaging is denoted by $\langle \rangle$. The F-2DES spectra of LH2 were measured and calculated with all-parallel pulse polarizations and thus contains a weak anisotropy contribution. Here, to isolate the population dynamics only, we will consider the isotropic signal acquired under the so-called magic angle between the polarization of the first and second pulse pairs, for which $\langle \mu_i \mu_i \mu_j \mu_j \rangle_{\text{m.a.}} = \frac{1}{9} |\mu_i|^2 |\mu_j|^2$. There are two limits under which the SE/GSB ratio reduces to the combinatoric result of Bolzonello et al. (29): a uniform population of all states regardless of the initial state (e.g., iso-energetic landscape), $U_{ji}(t_2) = \frac{1}{N} \forall i,j$, and all states with the same oscillator strength $|\mu_i|^2 = |\mu|^2 \forall_i$. For these cases, Eq. (1) for the ratio reduces to

$$\left(\frac{\text{SE}}{\text{GSB}}\right)_{\text{uniform population}} = \left(\frac{\text{SE}}{\text{GSB}}\right)_{\text{same oscillator strength}} = \frac{1}{N}. \quad (2)$$

However, none of these situations strictly applies to real aggregates. An interesting case is that of (quasi)equilibration within the excitonic manifold, for which we have $U_{ji}(t_2) = P_j^{eq} \forall i$. In this case, the $\sum_i |\mu_i|^2$ can be factored out and the SE/GSB ratio becomes

$$\left(\frac{\text{SE}}{\text{GSB}}\right)_{\text{equilibrated}} = \frac{\sum_{j=1}^{N} |\mu_j|^2 P_j^{eq}}{\sum_{j=1}^{N} |\mu_j|^2}. \quad (3)$$

In the ratio, the sum of the oscillator strengths (GSB) is compared to the same sum weighted by the populations. At thermal equilibrium at temperature $T$, the population of the excitonic states are $P_j^{eq} = \exp\left(-\epsilon_j/k_B T\right) \left(\sum_{k=1}^{N} \exp\left(-\epsilon_k/k_B T\right)\right)^{-1}$ where $\epsilon_j$ are the state energies. Depending on the excitonic state structure (energies, oscillator strength, delocalization), the SE/GSB ratio can thus be larger or smaller than 1/$N$.

Eqs. (1)-(3) are very simple and connect the static properties of the excitonic manifolds, without considering dynamics or line shapes, to the SE/GSB ratio that is observable as the weak dynamics in F-2DES spectra. To verify that the extracted SE/GSB ratio can indeed be inferred from the excitonic structure in realistic systems, we calculated the F-2DES spectra of various aggregates, as described in the SI. In F-2DES, the SE can be distinguished from GSB either by its spectrum or kinetics. In excitonic manifolds with many states, this can prove to be very difficult due to large spectral overlap and small SE/GSB ratio. This is the case of the individual LH2 rings as well, both of which have broad, featureless spectra. Fortunately, the excitation transfer between the rings allows us to distinguish the excited-state dynamics from the constant GSB contribution. We thus also construct our test aggregates with two spectrally distinct manifolds. For aggregates of varying geometries, sizes and excitonic delocalization, we have extracted the SE/GSB ratios and compared them to the results of Eq. (3), finding excellent agreement (see *SI Appendix*). This indicates that it is indeed the static excitonic state structure that dictates the presence of excited-state dynamics. These expressions apply to any aggregate that features distinct excitonic manifolds with rapid intra-manifold equilibration and slow inter-manifold energy transfer. One extreme case would be a perfect J aggregate for which all oscillator strength is in the lowest-energy state, leading to $\left(\frac{\text{SE}}{\text{GSB}}\right)_{\text{J-agg}} \to 1$. The opposite extreme case is a perfect H aggregate in which only the highest-energy state is bright, leading to a vanishing SE in the excited-state equilibrium and $\left(\frac{SE}{GSB}\right)_{H-agg} \rightarrow 0$. In general, the ratio will be somewhere between these two extremes, depending on the aggregate geometry, electronic coupling, energetic disorder and the excitonic delocalization [see Fig. 4a)]. In the case of the linear J/H aggregate, the SE/GSB ratio monotonously increases/decreases with the excitonic coupling relative to energetic disorder, correlating with the excitonic delocalization. In these aggregates, the SE/GSB ratio thus directly reports on the excitonic delocalization. A slightly more complex example is a circular aggregate in which the dipole moments are oriented tangentially to the circle. In a large circle (small coupling) the dominating interaction is between nearest neighbors, and the aggregate has J-type character, with the SE/GSB ratio increasing with delocalization. When decreasing the circle size (increasing coupling), the ratio reaches its maximum and then starts to decrease, signifying the transition to predominantly H-type behavior.

Having discussed the general aggregates, we briefly return to LH2 to understand the role of excitonic coupling and delocalization in the observed dynamics. Fig. 4b) depicts the dependence of the SE/GSB ratio on the excitonic delocalization for the LH2 B850 ring, in which both the electronic coupling and energetic disorder are relatively strong, leading to a significant excitonic delocalization and more complex state structure (48). At the same time, the equilibration within the B850 manifold is much faster than the B800→B850 transfer, so that Eq. (3) applies. To include the population transfer, Eq. (3) can be augmented as

$$\left(\frac{SE}{GSB}\right)_{B800} = \frac{\sum_{k=1}^{9} P_k^{eq}\left|\mu_{B800_k}\right|^2}{\sum_{k=1}^{9}\left|\mu_{B800_k}\right|^2} e^{-k_{B800 \rightarrow B850}t_2} \quad (4)$$

for the B800 ring, and

$$\left(\frac{SE}{GSB}\right)_{B850} = \frac{\sum_{j=1}^{18} P_j^{eq}\left|\mu_{B850_j}\right|^2}{\sum_{j=1}^{18}\left|\mu_{B850_j}\right|^2} (1 - e^{-k_{B800 \rightarrow B850}t_2}). \quad (5)$$

for the B850 ring. These expressions formalize the connection to the UDP decay and LCP rise observed in the experimental F-2DES data.

The geometry of the LH2 B850 18-*BChl a* ring is similar to that of the circular aggregate, leading to an analogous non-monotonous SE/GSB ratio dependence. For realistic LH2 parameters, the electronic nearest-neighbor coupling and energetic disorder are of a very similar magnitude which results in the lowest-energy states carrying most of the oscillator strength. This places the SE/GSB ratio close to its maximum in Fig. 4b). We thus have from just the excitonic structure, the rise of the LCP is $SE_{B850}/GSB_{B850}$ = 17% (18% for all-parallel), which is ~3 times larger than $1/N$=1/18 limit (5.5%) as well as the extracted experimental value of 6.2%. In the B800 ring the coupling and thus delocalization is much weaker, giving $SE_{B800}/GSB_{B800}$ = 13% (11% for all-parallel pulse polarizations). This matches the $1/N$=1/9 (11%) value for independent sites, but is again larger than the experimental value of 4.4%. As in the experiment, the SE/GSB ratio is about 1.5 times larger in the B850 ring (seen in the LCP) than in the B800 ring (seen in the UDP). The values are, however, larger than the experimental ones, highlighting the importance of including realistic lineshapes and the effects of finite laser bandwidth. A summary of the various SE/GSB ratios extracted from the different calculations is given in *SI Appendix* Table S1 for easy comparison with the experimentally determined ratios.

**Discussion**

The growing popularity of action-detected 2D spectroscopies motivates an in-depth analysis of their potential for probing excited state dynamics in molecular systems. While F-2DES has revealed excited-state dynamics in molecular dyads (18-20) and Rb atoms (16), recent work (29) raises doubts about the extent to which excited-state dynamics are visible in larger molecular aggregates. Using LH2 as a model system, our experimental data clearly demonstrates that F-2DES of LH2 shows very little spectral evolution as a function of the waiting time, in stark contrast to C-2DES (see Fig. 1). In particular, the kinetics of energy transfer from B800 to B850 are difficult to discern in F-2DES, constituting a ~6.2% change in the signal level of the LCP, compared to ~100% in C-2DES. The rise in the LCP in the F-2DES data occurs with a time constant of 1345 ± 575 fs, consistent with previous reports of the B800-B850 energy transfer time (49) and with our C-2DES measurement of 825 ± 10 fs. The substantially smaller relative kinetic signals in F-2DES compared to C-2DES means that high signal averaging and a stable experimental setup are required for resolving LH2 energy transfer kinetics with F-2DES. For observing energy transfer in LH2, C-2DES offers a clear advantage over F-2DES. However, the strength of the excited-state signatures in F-2DES carry important information about the excitonic structure as discussed below.

In order to understand the origin of the faint kinetic signatures in the F-DES spectra of LH2, we first examined the LH2 signals in the context of the combinatoric argument presented by Bolzonello et al. (29). At first glance, the experimentally extracted $SE_{B850}/GSB_{B850}$ of 6.2% appears in good agreement with the estimate of 5.6% obtained from the $1/N$ limit for the LCP. This fortuitous agreement incorrectly suggests an absence of excitonic delocalization in the B850 band. One would expect even better agreement with the $1/N$ limit for the B800 band due to its considerably weaker coupling in comparison to B850. Here we find that the $1/N$ limit fails, predicting $SE_{B800}/GSB_{B800}$ = 11% for the UDP, greatly exceeding the experimental value of 4.4%. Seeking better agreement with the experimental data we simulated the F-2DES spectra based on an excitonic model for LH2, using parameters consistent with previous work. Our F-2DES simulations showed weaker dynamic signatures than expected based strictly on the excitonic model of LH2. We attributed this to the insufficient separation of the broad B800 and B850 peaks in the F-2DES spectrum, and the effects of finite laser bandwidth. Taking both the laser spectrum and peak overlap into account we were able to obtain SE/GSB ratios in good agreement with the F-2DES measurements. The SE/GSB value reports on the interplay of excitonic coupling and disorder, allowing us to infer the delocalization within the B850 ring (Figure 3c) to be about 3.5 to 4 *BChl*s, in agreement with previous studies (42, 47, 50).

Through simulation we explored other molecular aggregate geometries, showing that the relative prominence of the excited-state dynamics is determined by the structure of the excitonic state manifold. Dependent on the aggregate geometry, the relative contribution of the excited-state dynamics can be greatly enhanced or suppressed. When the geometry of the system is known, there is a clear relation between the SE/GSB ratio and excitonic delocalization. This allows excitonic delocalization to be inferred within dense excitonic manifolds, which is not possible by standard C-2DES. While it might be difficult to distinguish the SE from GSB within the manifold by itself, the presence of an additional, spectrally separated manifold connected by slow excitation transfer is sufficient to make the SE visible. We note that future experimental advances, such as time-gating the fluorescence signal in F-2DES (30, 51) could allow for isolation of the SE signals, as could the recently-proposed 2D-FLEX method (52).

Given the difficulty in observing excited state dynamics in multichromophoric systems with F-2DES, the question arises as to what, if any, systems and questions are better-suited to action-detected spectroscopies compared to their coherent counterparts. Photocurrent-based 2DES measurements of solar cell materials have shown wide variability in their sensitivity to excited state dynamics between semiconductor-nanocrystal (8), perovskite (33) and organic samples (9). The work of Bolzonello et al. (29) suggests that the ability of action-based 2DES to resolve excited state dynamics decreases with system size, as predicted by their $1/N$ limit. They point out that "$N$ should be identified with the number of absorbing states rather than the number of independent chromophores". Eq. (3) is consistent with this idea, showing that the excitonic structure reweights the relative contributions of the states, showing that in the case of strong delocalization, it is possible to significantly beat the $1/N$ limit. Characterizing the SE/GSB ratio provides direct information about the delocalization and can be used to infer the excitonic structure. Remarkably, this works for dense excitonic manifolds as well, as long as the SE can be distinguished, for example by its dynamics as in LH2. An interesting effect occurs if large multichromophoric systems are connected to a smaller system containing only a few pigments. Our analysis suggests that energy transfer to the smaller system might be still observable in the cross-peak dynamics since the SE/GSB ratio depends only on the structure of final excitonic manifold [see Eq. (3)]. One particular system might be the Heliobacterial reaction center in which the antenna and the core of the reaction center can be spectrally separated (53). In addition, probing the excitonic structure via F-2DES might be particularly informative for systems with broad ESA signals such as photosynthetic reaction centers. Separated from their antenna complexes, these systems contain a small number of pigments. Understanding their excitonic structure in the low energy $Q_y$ region is an outstanding challenge due to a combination of disorder and spectral overlap of the constituent pigments. Attempts to leverage distinct $Q_x$ signatures of the pigments to untangle the $Q_y$ excitonic structure have been complicated by the broadband ESA signals in the $Q_x$ region (54). Should the oppositely-signed ESA pathways in F-2DES cancel in such systems as they do in LH2, broadband F-2DES measurements may offer advantages over C-2DES for revealing excitonic structure.

**Materials and Methods**

The LH2 samples from *Rbl. acidophilus* were prepared following established protocols (55) that are described in detail in the *SI Appendix.* Our experimental setup for the F-2DES measurements is based on the acousto-optic phase modulation and lock-in detection scheme first demonstrated by Marcus and co-workers (5, 56), and used previously by us (6, 20). We describe our current implementation in detail in the *SI Appendix*. The pulse energies used were 5.5 pJ per pulse with spot sizes of 16 μm, exciting ~10% of the sample at each laser shot. Samples were flowed at ~75 mL/min to reduce photobleaching effects. The $t_1$ and $t_3$ delays were scanned from 0 to 91 fs, in steps of 7 fs, while $t_2$ was scanned from 0 to 2000 fs. A lock-in time constant of 100 ms was used in the measurements and the signal at each combination of the three delays was averaged over 60 acquisitions (8 min per $t_2$ delay). The C-2DES measurements were made in the pump-probe geometry as reported previously (57). The pump pulses had a total pulse energy of 7.3 nJ and were focused to 190 μm at the sample position, whereas the probe pulse had a pulse energy of 2.5 nJ and was focused to 180 μm. The $t_1$ delay was scanned from 0 to 180 fs, in steps of 5 fs, in the partially rotating frame with 0-π phase cycling. For the B800-B850 energy transfer kinetics, $t_2$ was scanned from -100 to 5000 fs in 100 fs steps. Each $t_2$ spectrum was collected in ~1.5 minutes, requiring ~90,000 total laser shots. Further details on the C-2DES implementation are in the *SI Appendix*.

The excitonic model of LH2 was constructed as follows (additional details in the *SI Appendix*). Using the structure from the RSCB protein databank [2FKW] (58), we calculate the electronic coupling via the dipole-dipole approximation. We assume identical site energies for the B800 band, and *BChl a* dimers in the B850 band (59), with Gaussian energetic disorder three-times stronger in the B850 band, reflecting possible charge-transfer character (60). The static excitonic structure is depicted in Fig. 3a. The excitonic delocalization can be calculated as $l_{\text{deloc}} = N \left( \sum_{i,n=1}^{N} |c_n^i|^4 \right)^{-1}$, where $c_n^i$ are the coefficients of the transformation from the site basis (index *n*) to the excitonic basis (index *i*), where *N*=9 for the B800 ring and N=18 for the B850 ring. We describe the vibrational bath using a 3-component Brownian oscillator model, and intra-manifold population relaxation rates as well as coherence dephasing are calculated using Redfield theory. The excitonic lineshapes are calculated by a cumulant expansion using the same spectral density, considering exchange-narrowing and relaxation-induced broadening. For faster calculations in the frequency domain, the lineshape was fit to a Lorentzian. The waiting time propagator $U(t_2)$ was calculated using the secular approximation and matrix exponential, with the excitation transfer between the rings described by a site-independent rate with time constant of 830 fs.

**Acknowledgments**

We gratefully acknowledge the support of the National Science Foundation through Grant #PHY-1914608 (S.E.S. and J.P.O), the AFOSR Biophysics program under Grant No FA9550-18-1-0343 (A.J.) and Grant No FA9550-21-1-0098 (J.L.). R.W. and M.Z. gratefully acknowledge the support of the Office of Basic Energy Sciences, the U.S. Department of Energy Grant DE-SC0016384. J.L. acknowledges support from the HFSP fellowship program under Grant No LT0056/2024-C.

**Figures and Tables**

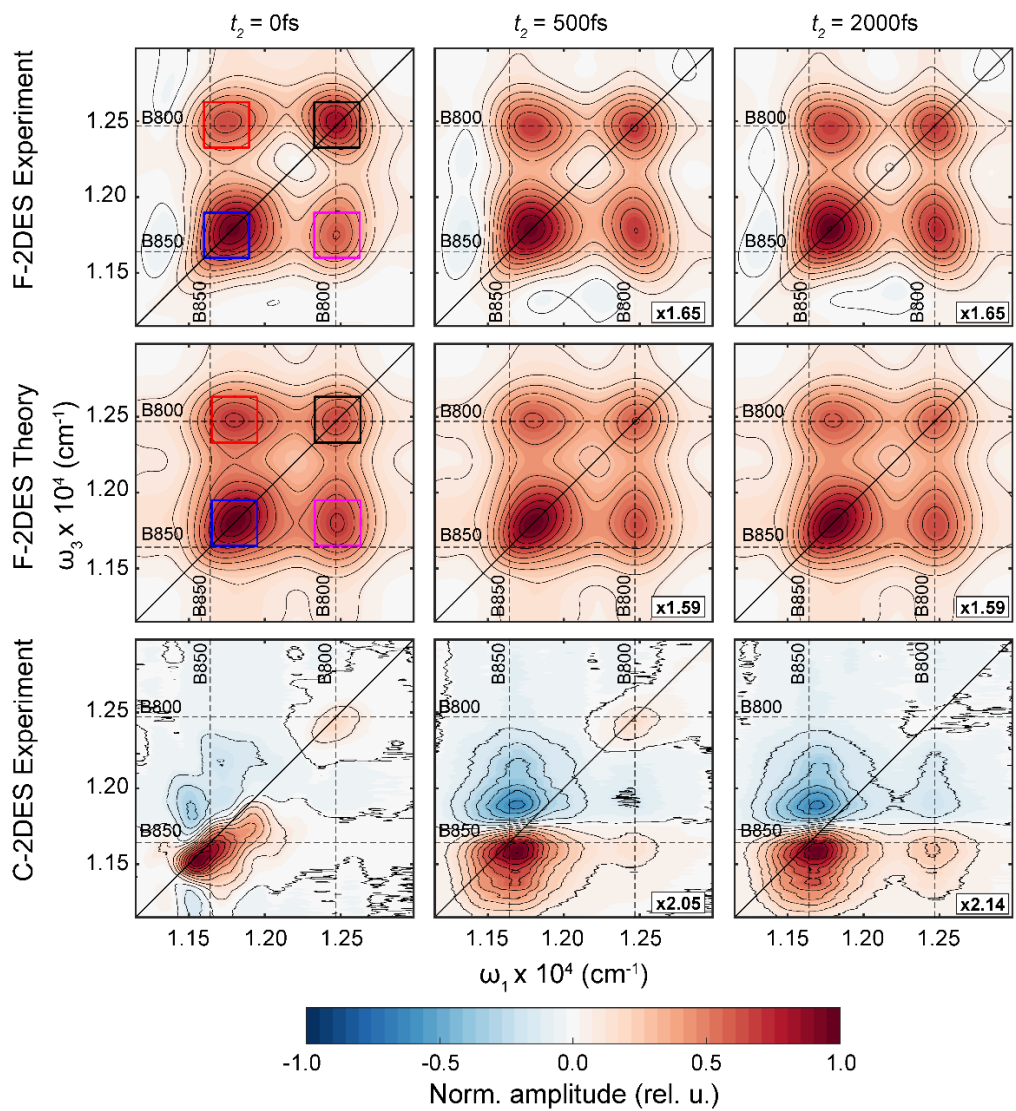

**Fig. 1.** Top row: F-2DES spectra of LH2 at waiting times $t_2$=0 fs, 500 fs and 2 ps. Middle row: Simulated F-2DES spectra of LH2 at waiting times $t_2$=0 fs, 500 fs and 2 ps. Bottom row: C-2DES spectra of LH2 at waiting times $t_2$=0 fs, 500 fs and 2 ps. The colored squares in the $t_2$=0 fs spectra in the top and middle row indicate the region over which the kinetic traces shown in Fig. 2 were averaged.

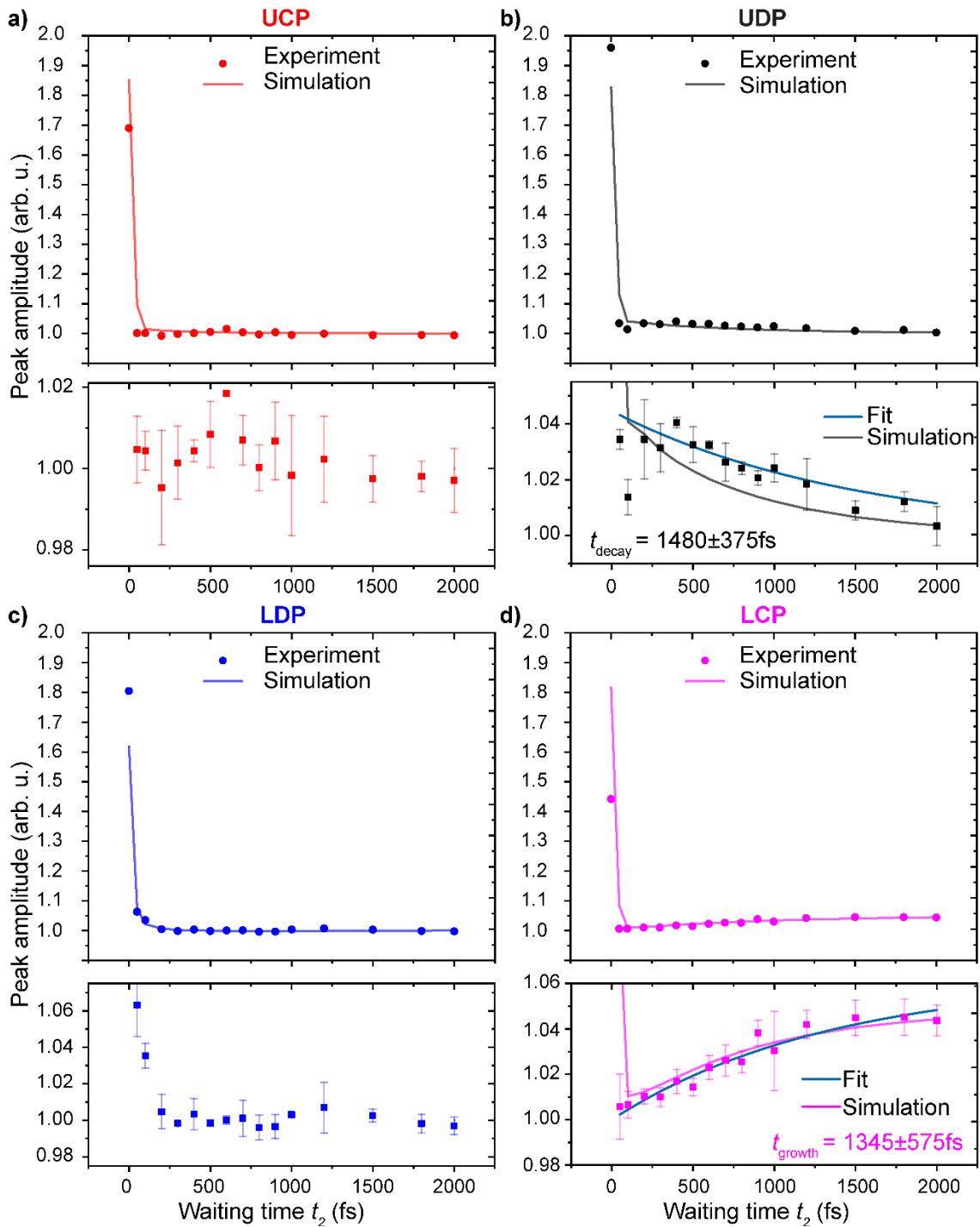

**Fig. 2.** Experimental and simulated kinetic traces as a function of waiting time $t_2$ at a) the upper cross peak (UCP), b) upper diagonal peak (UDP), c) lower diagonal peak (LDP) and d) lower cross peak (LCP) locations, obtained by averaging over the regions indicated by colored squares in Fig. 1. Below each kinetic trace, an expanded vertical scale excludes the $t_2$ = 0 fs peak to more clearly visualize the weak kinetic signatures. Error bars are derived from three distinct measurements. Exponential fits to the UDP and lower cross-peak LCP are also shown, yielding $SE_{B800}/GSB_{B800}$ = 4.4% and $SE_{B850}/GSB_{B850}$ = 6.2% respectively. The UCP and LDP were normalized such that the peak amplitude at long waiting times (>1ps) was 1 on average. The traces were normalized to the GSB contribution as further described in the *SI Appendix.*

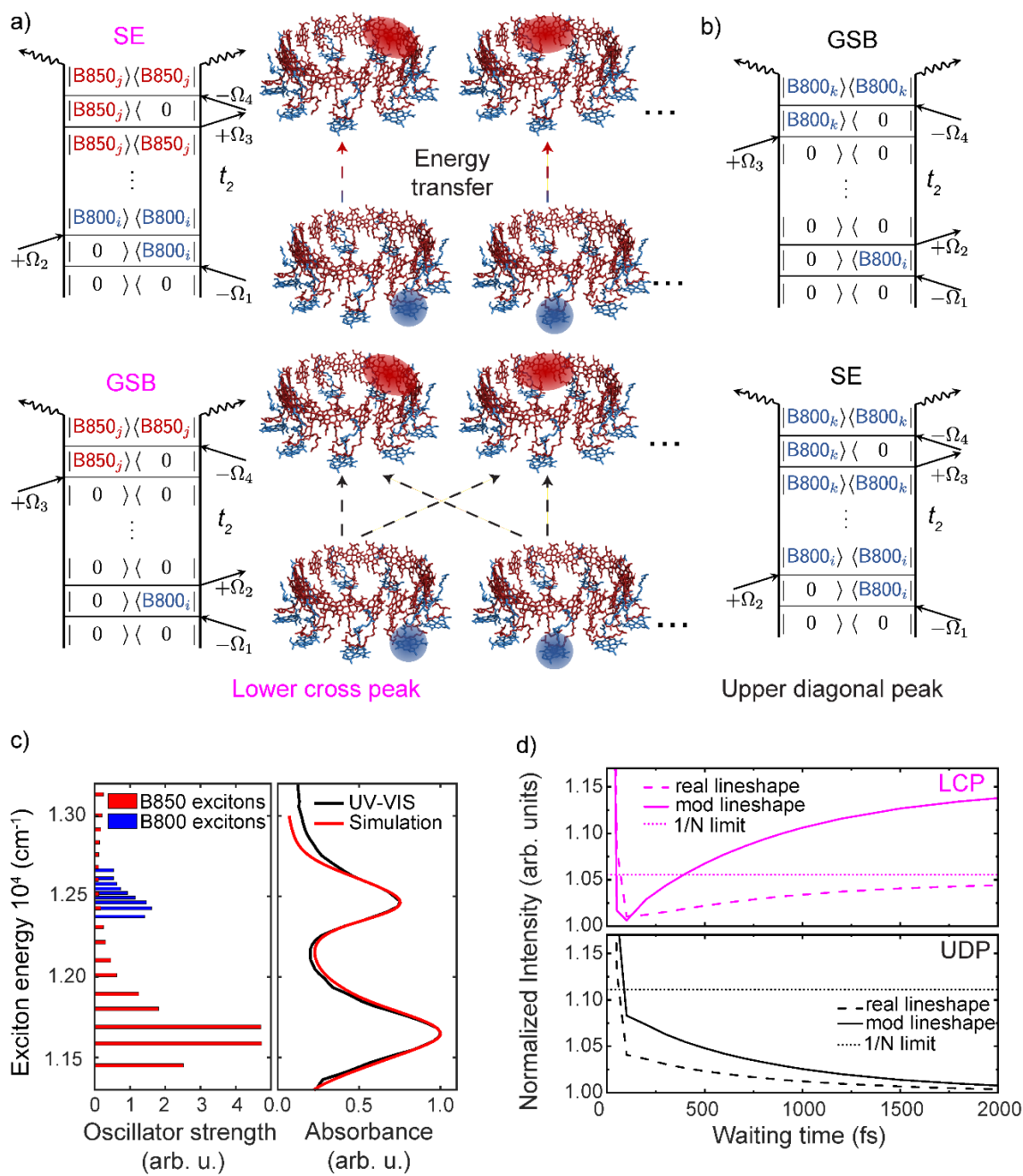

**Fig. 3** a) Double-sided Feynman diagrams depicting the signals that contribute to the lower cross peak (LCP) which exhibits B800 to B850 energy transfer (ET) during waiting time $t_2$. The structure was taken from the RSCB protein databank [2FKW]. (58) b) Double-sided Feynman diagrams depicting the signals that contribute to the upper diagonal peak (from B800). c) Excitonic model of LH2, showing the energies and oscillator strengths of the excitonic states of the B800 and B850 bands (left). Experimental (black) and simulated (red) linear absorption spectrum of LH2 (right). d) Simulated $t_2$-dependent kinetic traces of the lower cross peak (upper panel) and upper diagonal peak (lower panel) using realistic lineshapes (solid) and modified lineshapes that take into account the laser spectrum (dashed). The traces were normalized to the GSB contribution as further described in the *SI Appendix*. The 1/*N* limits are plotted at the 1/*N* value above 1 for the LCP/UDP.

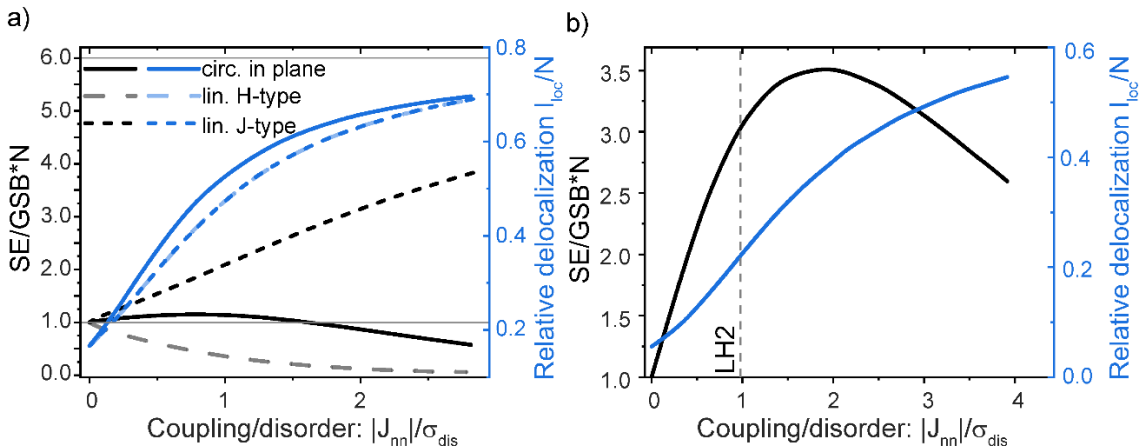

**Fig. 4.** a) Dependence of the SE/GSB ratio on the excitonic delocalization for different types of molecular aggregates. Here the electronic coupling is characterized by the nearest-neighbour coupling $J_{nn}$, while $\sigma_{dis}$ characterizes the Gaussian disorder (mean square deviation). The delocalization length is calculated as described in the Methods section. Thin, gray lines at 1 and 6 are the 1/*N* value for identical molecules and the maximum ratio for a perfect J-aggregate with 6 monomers. b) Dependence of the SE/GSB ratio on the excitonic delocalization for the LH2 B850 ring. The dotted line indicates the $|J_{nn}|/\sigma_{dis}$ used in the F-2DES simulations, yielding a SE/GSB ratio of ~3 times larger than the 1/*N* limit.

## Supporting Information Appendix:

## Photosynthetic energy transfer: missing in action (detected spectroscopy)?

Ariba Javed[1,2†], Julian Lüttig[1†], Kateřina Charvátová[3], Stephanie E. Sanders[1], Rhiannon Willow[1], Muyi Zhang[1], Alastair T. Gardiner[4], Pavel Malý[3*] and Jennifer P. Ogilvie[1*]

[1]Department of Physics, University of Michigan, 450 Church St, Ann Arbor MI 48109, USA

[2]Department of Materials Science and Engineering, University of Michigan, 2300 Hayward St., Ann Arbor, Michigan 48109-2136, USA

[3]Faculty of Mathematics and Physics, Charles University, Ke Karlovu 5, 121 16 Praha 2, Czech Republic

[4]Center Algatech, Institute of Microbiology, Czech Academy of Sciences, 37901 Třeboň, Czech Republic

† Equal author contributions

*Corresponding authors: Jennifer P. Ogilvie (experiment) and Pavel Malý (theory)

**Fluorescence-detected 2DES: Double-sided Feynman Diagrams**

In Fig. S1 and S2 we show the double-sided Feynman diagrams representing the rephasing signal of the upper diagonal peak (B800-B800, UDP) and the lower cross peak (B800-B850, LCP) in F-2DES. The signal is selected via its specific phase modulation frequency given by $+\Omega_1 - \Omega_2 + \Omega_3 - \Omega_4$, as described in the F-2DES experimental methods section below. Note that we only show explicitly the rephasing diagrams. However, our discussion of the rephasing diagrams applies analogously to the non-rephasing diagrams, which possess a different phase signature. In contrast to Fig. 3 in the main manuscript, where we only discuss the weights of the GSB and SE pathways in detail, here we also show the two ESA pathways, i.e. ESA1 and ESA2. Note that both ESA1 and ESA2 have two possible pathways due to the two rings in LH2 (middle and bottom rows of Fig. S1 and S2). Two of the four total ESA pathways involve a doubly excited state on a single ring ($|B800_\alpha\rangle$ and $|B850_\beta\rangle$, Fig. S1 and S2, middle row) for the UDP and LCP respectively. The lower rows of Fig. S1 and S2 depict double-sided Feynman diagrams with a "mixed" doubly excited state where the two excitation are located on two different rings. Both ESA1 and ESA2 include the same dipole moments but differ in their final population: ESA1 ends in a population of a singly excited state, while ESA2 ends in a population of a doubly excited state. Here we follow the convention in literature (2) where pathways with an even number of interactions from the right have a positive sign (GSB, SE and ESA1), while pathways with an even number of interactions from the left are considered with a negative sign (ESA2). The sign of each pathway is indicated underneath the corresponding pathway. As thoroughly discussed in literature (3-5) the yield of the doubly excited state would be twice as much as that of the singly excited state. However, the yield of the ESA2 pathways can be reduced either by sample properties such as exciton-exciton annihilation or detection effects such as incoherent mixing (3, 5-7). In the case of LH2 we expect to have efficient annihilation resulting in the same yield of ESA1 and ESA2 pathways and complete cancellation of both pathways.

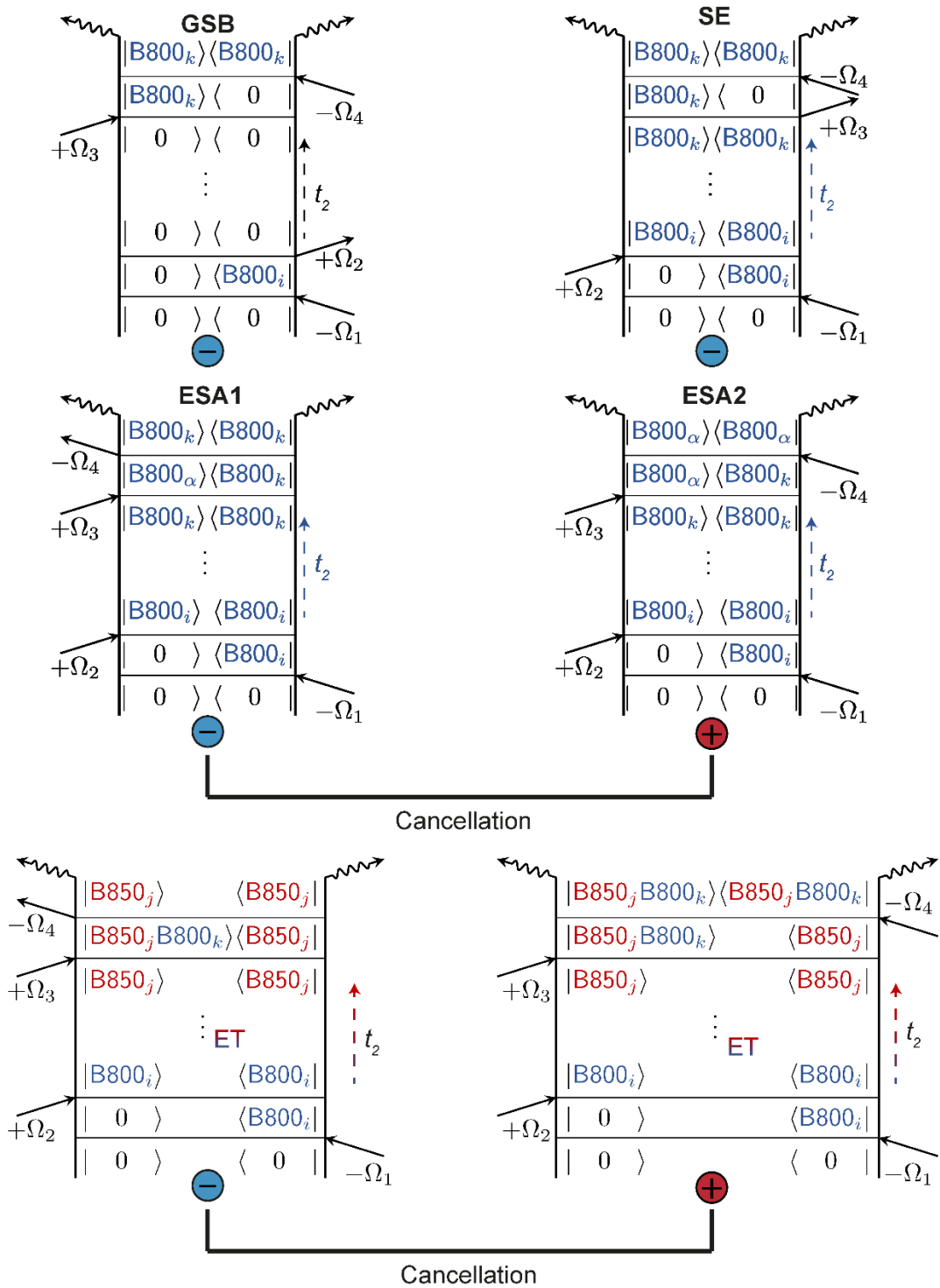

**Figure S1.** Double-sided Feynman diagrams representing the rephasing signal of the upper diagonal peak (B800-B800) in F-2DES. The sign of each pathway is indicated underneath each diagram. For ESA1 and ESA2, there are two possible pathways.

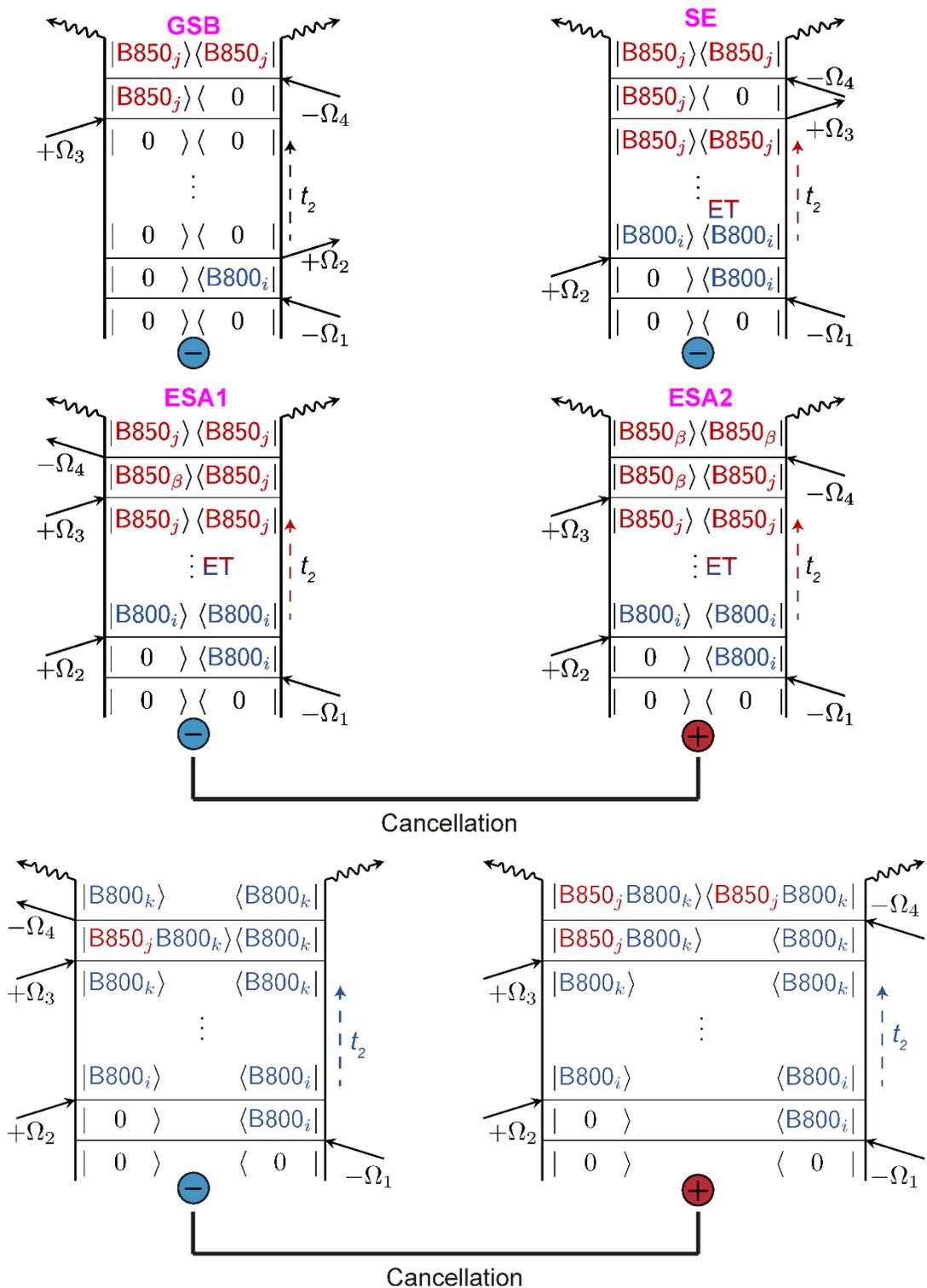

**Figure S2.** Double-sided Feynman diagrams representing the rephasing signal of the lower cross peak (B800-B850) in F-2DES. The sign of each pathway is indicated underneath each diagram. For ESA1 and ESA2, there are two possible pathways.

## Coherent-detected 2DES: Double-sided Feynman Diagrams

In contrast to F-2DES, the signal in C-2DES is connected to the nonlinear polarization of the sample after interaction with three electric fields. Let us consider the dynamics of energy transfer in C-2DES of LH2 in the weakly coupled limit. We are here only showing rephasing pathways in the phase matching condition of $-\mathbf{k}_1 + \mathbf{k}_2 + \mathbf{k}_3$. Additionally, the nonrephasing pathways are emitted in the corresponding phase matching condition and the argument below applies to them as well. The Feynman diagrams for the UDP and LCP of LH2 in C2DES are shown in Fig. S3 and Fig. S4. For the UDP four pathways must be considered: a GSB pathway, a SE pathway, and ESA pathways that can be further distinguished if intra-ring (left) or inter-ring (right) transport occurs during the waiting time. The former pathway ("intra-ring") involves the doubly excited state ($|B800_\alpha\rangle$) in the B800 ring. The latter pathway ("inter-ring") can be understood as an excitation of B800 with transfer to B850 and subsequently probing of the B800 states. The two ESA pathways are both negative. Therefore, they do not cancel in C-2DES, and contribute to the UDP (Fig. S3) and LCP (Fig. S4). In contrast to F-2DES, energy transfer occurs during the waiting time of the SE as well as the ESA pathways and the interplay between the two pathways lead to an amplitude decrease of the UDP. The yield of each pathway is the same since in C-2DES the coherent field is detected rather than a signal that is proportional to the excited state population as in F-2DES.

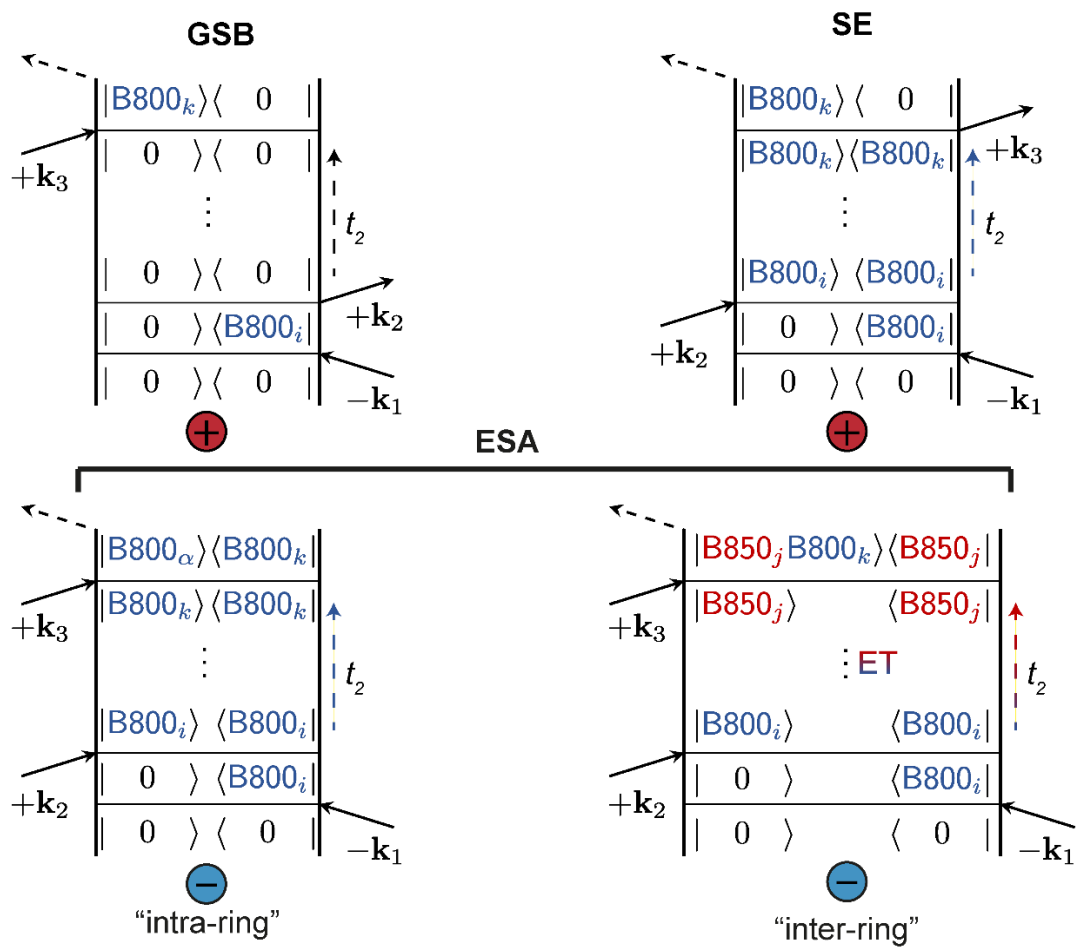

**Figure S3.** Double-sided Feynman diagrams for the rephasing signal in C-2DES for the upper diagonal peak (B800-B800).

For the GSB pathway the indices *i* and *k* can be any pigment in B800 resulting in a combinatoric weight of $N_{B800} \cdot N_{B800}$. The stimulated emission is only possible if we probe the excited pigment reducing the combinatoric weight to $N_{B800}$. The "intra-ring" ESA pathway includes a doubly excited state of B800 ($|B800_\alpha\rangle$). After the first interaction with $N_{B800}$ possible pigments, we only count the combinations where a pigment other than *i* is excited because exciting the same pigment twice might contribute to another spectral position in the 2D map. Therefore, the resulting total weight of this ESA pathway is $N_{B800} \cdot (N_{B800} - 1)$. The "inter-ring" ESA pathway has a combinatoric weight of $N_{B800} \cdot N_{B800}$ since the initial excitation in B800 is transferred to B850 during the waiting time leaving again $N_{B800}$ combinations for the interaction with the probe. Energy transfer is represented by propagators during the waiting time. For the SE and "intra-ring" ESA pathways this propagator is $e^{-k_{B800 \to B850} t_2}$, i.e., a decay with energy transfer, while the "inter-ring" ESA pathway will increase with $1 - e^{-k_{B800 \to B850} t_2}$. We can summarize the amplitude of each pathway over the waiting time $t_2$ as

$$\text{GSB} \propto N_{B800} \cdot N_{B800} \tag{S1}$$

$$\text{SE} \propto N_{B800} \cdot (e^{-k_{B800 \to B850} t_2}) \tag{S2}$$

$$\text{ESA} \propto \underbrace{-N_{B800} \cdot (N_{B800} - 1)(e^{-k_{B800 \to B850} t_2})}_{\text{"intra-ring"}} \underbrace{-N_{B800} \cdot N_{B800}(1 - e^{-k_{B800 \to B850} t_2})}_{\text{"inter-ring"}} \tag{S3}$$

where the proportionality sign reflects the fact that we did not include specific lineshape functions and the explicit prefactor of transition dipole moments. However, the transition dipole moments are the same for all four pathways, so we omit them here for simplicity. At $t_2 = 0$ only the GSB, SE and "intra-ring" ESA pathways contribute to the signal and the "intra-ring" ESA pathway has an opposite sign compared to the GSB and SE pathways. However, the total combinatoric weight of GSB and SE is larger than the one of the "intra-ring" ESA pathway, resulting in partial cancellation. This is in accordance with the observation of a non-zero signal for the diagonal peak at $t_2 = 0$. Additionally, the frequency of the ESA can be slightly shifted compared to the GSB and SE, which results in the characteristic derivative lineshape of C-2DES spectra. For large waiting times the amplitudes of the SE and "intra-ring" ESA pathways have decreased completely and the amplitude of the ESA pathway with inter-ring transport has increased to its maximum. Since the GSB and "inter-ring" ESA pathways have opposite sign, this leads to an overall reduction of the signal amplitude with the energy transfer rate. For $t_2 \gg 0$, the GSB and "inter-ring" ESA pathways will cancel completely since they have the same combinatoric weight but opposite sign leading to a diagonal peak equal to zero after efficient transfer occurred.

For the LCP the rephasing signal can also be described by GSB, SE, and ESA pathways. The ESA pathways can again be separated into "inter-ring" and "intra-ring" pathways. The combinatoric weight for the GSB and "intra-ring" ESA pathways is $N_{B850} \cdot N_{B800}$. The "inter-ring" ESA pathway has the weight of $N_{B800} \cdot (N_{B850} - 1)$ since the diagrams where the same pigment is excited twice are not considered. The SE pathway has a combinatoric weight of $N_{B800}$ since for this diagram the probe can only interact with the B850 pigment the exciton transferred to. The SE and "inter-ring" ESA signals will rise with $1 - e^{-k_{B800 \to B850} t_2}$, while the "intra-ring" ESA pathway decays with $e^{-k_{B800 \to B850} t_2}$. In summary the amplitudes of the different pathways are

$$\text{GSB} \propto N_{B800} \cdot N_{B850} \tag{S4}$$

$$\text{SE} \propto N_{B800} \cdot (1 - e^{-k_{B800 \to B850} t_2}) \tag{S5}$$

$$\text{ESA} \propto \underbrace{-N_{B800} \cdot (N_{B850} - 1) \cdot (1 - e^{-k_{B800 \to B850} t_2})}_{\text{"inter-ring"}} \underbrace{-N_{B800} \cdot N_{B850} \cdot e^{-k_{B800 \to B850} t_2}}_{\text{"intra-ring"}} \tag{S6}$$

At $t_2 = 0$, only the GSB and "intra-ring" ESA pathways are nonzero. Since the two pathways have opposite signs but the same combinatoric weight, they compensate for each other, and the net signal is zero. This fact is in accordance with the observation that the LCP in C-2DES is not visible before energy transfer has occurred, i.e., at $t_2 = 0$. Note that for strong delocalization a non-zero cross peak at $t_2 = 0$ would be present due to the redistribution of oscillator strength (3). With the energy transfer rate, the amplitude of the "intra-ring" ESA pathway will decrease and the amplitude of SE and "inter-ring" ESA pathways will rise breaking the cancellation of pathways at $t_2 = 0$. The result is an overall rise of the cross peak from zero. The observed derivative lineshape is caused by slight frequency differences of the SE and "inter-ring" ESA pathways where excitation from ground state to the singly excited state and from the singly excited state to the doubly excited state differ in their specific frequencies.

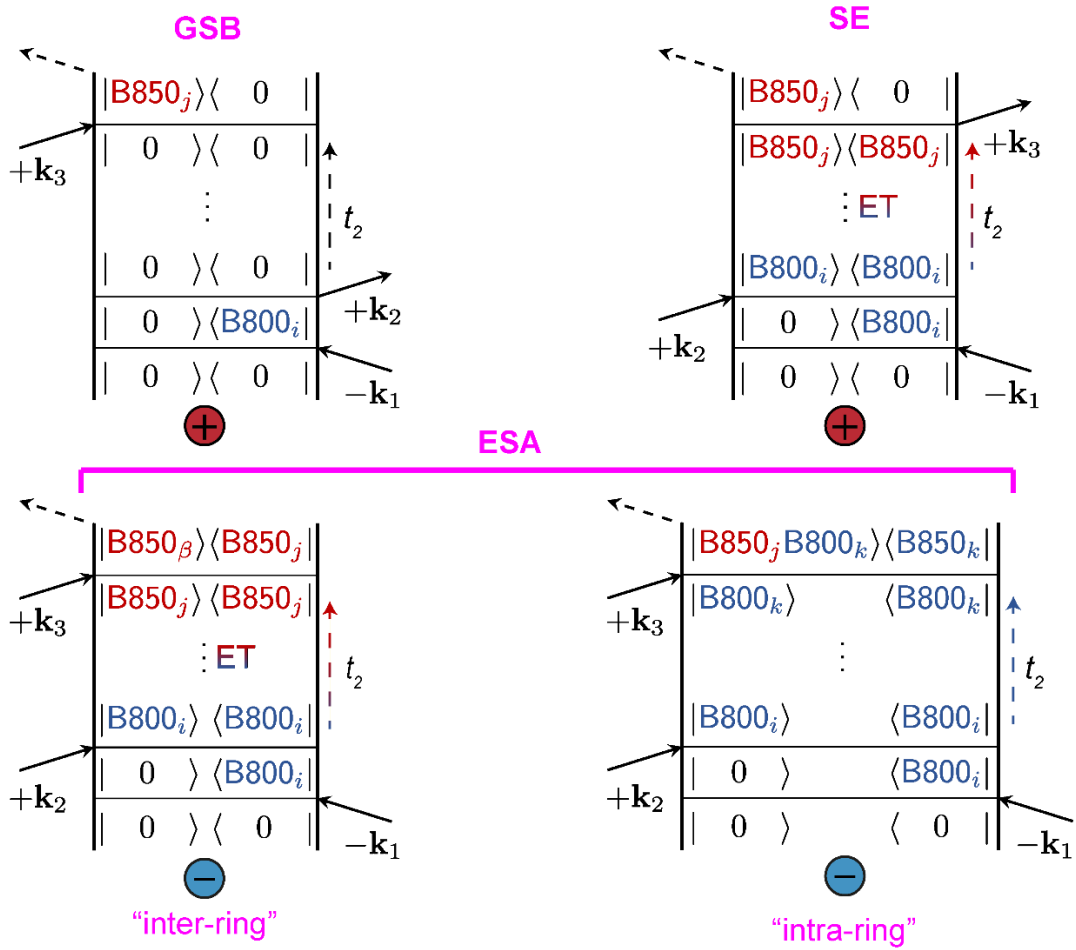

**Figure S4.** Double-sided Feynman diagrams for the rephasing signal in coherent 2DES for the lower cross peak (B850-B800).

**Sample Preparation**

Cells of *Rbl. acidophilus* strain 10050 were grown anaerobically in C-succinate media at 30 °C at a light intensity of ~ 100 µmol photons $s^{-1}$ $m^{-2}$. The fully grown culture was pelleted, washed once with 20 mM MES, 100 mM KCl, pH 6.8 to remove media traces and the cell pellet either used immediately or flash frozen until required. The *Rbl. acidophilus* cells were re-suspended in 20 mM Tris.Cl pH 8.0, homogenized thoroughly with a few grains of DNAse and a few mg of $MgCl_2$. The cells were broken by passage through an Emulsiflex-CS cell disrupter three times and the sample was immediately ultra-centrifuged (180,000 × g, 90 min, 4 °C) to pellet the membranes. The supernatant was discarded, and the membranes were gently re-suspended with 20 mM Tris.Cl pH 8.0 and adjusted to an optical density (OD) at the $Q_x$ maxima (approx. 580 nm) of 50 in a 1 cm pathlength cuvette.

The membrane suspension was solubilized at room temperature with gentle stirring for 1 h with 1% N, N-dimethyldodecylamine N-oxide (LDAO) and then centrifuged (100,000 x g, 30 min) to remove any un-solubilized material. The solubilized, supernatant fraction was fractionated using stepwise sucrose density centrifugation (150,000 × g, 4 °C, 16 h). The LH2 complex band was gently removed from the gradient, loaded on an anion-exchange gravity column with ToyoPearl 650S resin (Tosoh) and eluted with increasing concentrations of NaCl in TL buffer

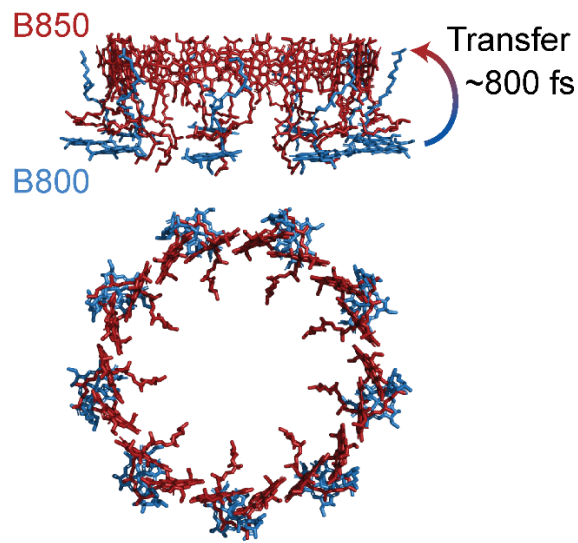

**Figure S5.** Crystal structure of LH2 showing the arrangement of bacteriochlorophyll pigments in the B800 and B850 rings. Energy transfer from B800 to B850 occurs on a timescale of roughly 800 fs. [Structure taken from protein data bank reference 2FKW (1)]

(20 mM Tris.Cl, 0.1% LDAO, pH 8.0). The resulting fractions were assayed spectrophotometrically for purity. Then, the best fractions (Abs. 858 nm / Abs 275 = 3.2) were pooled, run on a Sephacryl S-300 (GE Healthcare) size-exclusion column, concentrated and flash frozen until use. The crystal structure of the resulting LH2 complexes depicting only the BChl a molecules is shown in Fig. S5. In this publication we focus on energy transfer from the B800 subunit to the B850 subunit.

**Phase-Modulation Fluorescence-Detected 2DES Experimental Setup**

The experimental setup used for the F-2DES measurements is based on the acousto-optic phase modulation and lock-in detection scheme first demonstrated by Marcus and co-workers and recently implemented by us (8-10). Acousto-optic modulation enables the use of high modulation frequencies, resulting in the signals being modulated at different linear combinations of such high frequencies (in the kHz range), thus enabling physical undersampling of the signal and minimizing $1/f$ noise in the measurements. Lock-in detection provides the advantage of high sensitivity over a wide dynamic range and enables the selection of specific frequencies against a large noise background.

A layout of the setup is shown in Fig S6. The pulse train from a 1 MHz ytterbium amplified laser (Spectra Physics Spirit-HE, 1 MHz, 280 fs pulse duration) is focused into an 8 mm thick Yttrium Aluminum Garnet (YAG) crystal using a 5 cm focusing lens, resulting in white-light generation (WLG). Re-collimation of the beam to a 3.7 mm diameter ($1/e^2$) is obtained using a 4 cm focal length lens. A short pass (SP 950 nm OD4, Edmund optics) filter is used to reject the residual fundamental light. To ensure that the pulses are Fourier-transform-limited at the sample position, the beam is routed to a SLM (spatial light modulator)-based pulse shaper (MIIPS 640P, Bio-photonic Solutions) for dispersion pre-compensation and further bandwidth reduction. After the pulse shaper, a 50/50 beam splitter (BS1, Newport, 10B20BS.2) splits the beam equally and directs it to two Mach-Zehnder interferometers (MZ1 and MZ2) to generate the excitation and detection pulse pairs.

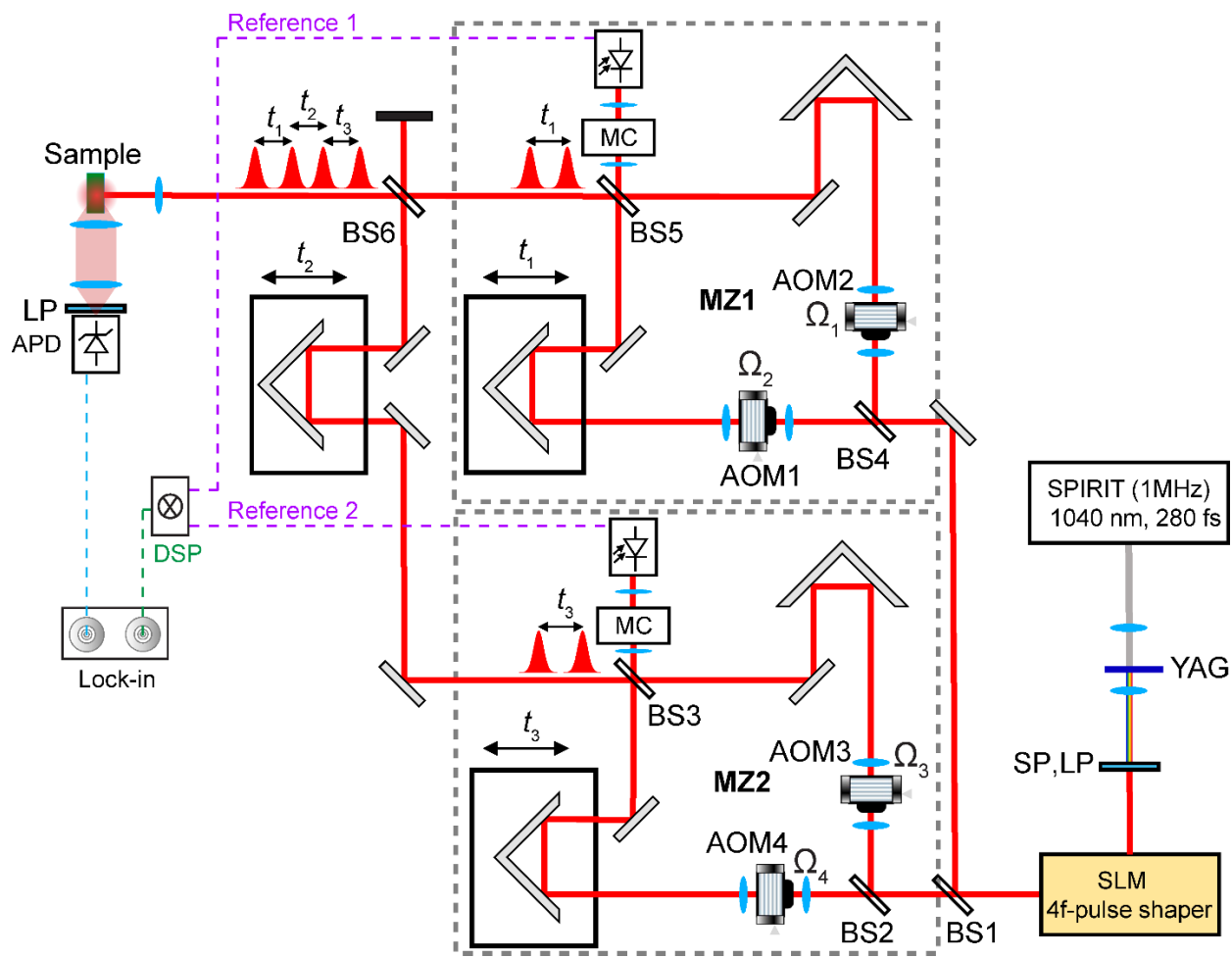

**Figure S6:** F-2DES setup based on phase modulation with lock-in detection.

At the entrance of each interferometer, beam splitters (BS2 and BS4) divide the pulse energy equally along the MZ's fixed and moving arms. An acousto-optic modulator (AOM, Isomet, M1142-SF80L) inserted in the optical path of each arm, modulating at a particular radio frequency $\Omega_i$, imparts a time-varying phase shift to the consecutive pulses of the pulse train (8). All four AOMs are driven by the same RF signal generator (Novatech, 409B), which uses a common internal clock for generating the modulation signals $\Omega_{1,2,3,4}$ = (80.110, 80.103, 80.000, 79.975) MHz. Each pair of the modulated pulses is recombined on a beam splitter (BS3 and BS5) at the interferometer outputs, with the interpulse delay within each interferometer ($t_1$ and $t_3$), varied using a mechanical delay stage (DS, Newport, M-VP25XL). While the time delays $t_1$ and $t_3$ correspond to the coherence and detection times, respectively, the waiting time $t_2$ is controlled by an additional delay stage (DS2, Newport, ILS150). All three mechanical stages are controlled using a motion controller (Newport, XPS-Q4). The recombined outputs from BS3 and BS5 oscillate at the difference frequency of the AOMs, that is, $\Omega_{1,2}$ = 7 kHz and $\Omega_{3,4}$ = 25 kHz respectively. One half of the outputs from BS3 and BS5 is routed to a monochromator (MC, Dynacil, MC1-05G, 50 μm slit) for spectral narrowing at a center wavelength $\omega_{Ref}$ = 825 nm with a FWHM bandwidth of 4 nm. The spectrally narrowed monochromator signals, Reference 1 and Reference 2 from MZ1 and MZ2 respectively are collected by variable-gain photodiodes (PD, Thorlabs PDA36A, set at 20 dB gain) and sent to a 24-bit digital signal processor (DSP, Analog Devices, ADAU1761). The DSP mixes the two monochromator signals Reference 1 and Reference 2 to generate the reference signals corresponding to rephasing and non-rephasing frequencies at $\Omega_R = -\Omega_{1,2} + \Omega_{3,4}$ = 18 kHz and $\Omega_{NR} = +\Omega_{1,2} + \Omega_{3,4}$ = 32 kHz respectively. The DSP applies appropriate filtering and gain to the reference signals. A lock-in amplifier (Zurich Instruments, HF2LI) is referenced to the DSP signals and is used to demodulate the fluorescence signals oscillating at $\Omega_R$ and $\Omega_{NR}$. The lock-in detection of the fluorescence signal is carried out relative to the reference wavelength, $\omega_{Ref}$, enabling physical undersampling (9) of the signal. This means that the signal centered at $\omega_{eg}$ corresponding to the electronic energy gap of the material can be sampled at a lower frequency $\omega_{eg} - \omega_{Ref}$, allowing for faster measurement times. In addition, undersampling makes the measurement insensitive to phase noise resulting from mechanical delay fluctuations. The other outputs from BS3 and BS5 are recombined at BS6 to form a collinear four-pulse train and is used to excite the sample. All four beams have parallel relative polarization. The unused output from BS6 is directed to a beam-dump.

Prior to the experiments, the pulses are compressed to 17 fs (FWHM) using the 4f-SLM pulse shaper in a 20 μm thick β-barium borate (BBO) crystal at the sample position. The four-pulse sequence interacts with the sample in the 90° detection scheme, where the four collinear pulses are focused with a lens of focal length 5 cm to a spot size of ~16 μm (FWHM) at the sample position. The fluorescence is collected in the 90° direction using a large aperture collimating lens (Thorlabs LA1401-AB, 2" dia) with a focal length of 6 cm, and then focused onto an avalanche photodiode (APD, Hamamatsu C12703-01) with a lens of focal length of 10 cm (Thorlabs LA1442-B, 2" dia). Longpass (865 nm and 887 nm) filters are used to prevent any laser scatter from entering the APD. The signal from the APD is split into two channels and sent to the lock-in amplifier for detection of rephasing and non-rephasing signals. Step-scanning of the time delays was used for the measurements reported in this work. At each $t_2$ delay, a 2D grid was recorded as a function of $t_1$ and $t_3$. In our collection scheme, $t_1$ takes the shortest time to measure while $t_2$ delay takes the longest. At $t_1, t_2, t_3 = 0$, the 2D signal is phased in the time domain by the lock-in amplifier. A lock-in time constant of 100 ms was used in the measurements and the signal at each combination of the three delays $t_1, t_2, t_3$ delays was averaged over 60 acquisitions. The $t_1$ and $t_3$ delays were scanned from 0 to 91 fs, in steps of 7 fs, while $t_2$ was scanned from 0 to 2000 fs. The sample was continuously flowed at ~75 mL/min through a 1 mm pathlength cuvette with a peristaltic pump to reduce the effect of photobleaching. The OD used in the measurements was ~0.35 in a 1 mm pathlength.

**Phase-Modulation F-2DES lock-in data analysis**

Real and imaginary values of rephasing and non-rephasing data were output using lock-in detection for all combinations of $t_1, t_2, t_3$ delays. First of all, an initial phase removal was performed on the entire 3D dataset using the phase of the $t_1, t_2, t_3 = 0$ data point. The time domain data at each $t_2$ were then asymmetrically zeropadded after 91 fs, to a total of 256 data points along both $t_1$ and $t_3$. A 2D apodization function was then applied to the time domain data to bring the data smoothly to zero at the positive edges. This 2D window function was obtained by the multiplication of identical inverse tangent filter functions in the $t_1$ and

$t_3$ dimensions, given by the form $w(\tau) = \frac{1}{2}\{1 - \tanh\left[\frac{\tau-\tau_0}{\sigma}\right]\}$, where $\tau_0 = 84$ fs and $\sigma = 7$ fs. Data along $t_1, t_3 = 0$ were set to half the amplitude according to the half-maximum convention (11). This step yields a frequency domain 2D spectrum without a background offset in the shape of a "cross". Double Fourier transformation of the data was then performed with respect to $t_1$ and $t_3$ to obtain a 2D rephasing or non-rephasing spectrum at each $t_2$. Rephasing spectra were flipped along the $t_1$ axis and a circular shift operation was used to account for the shift of one index in case of even number of data points. (256 in our case). Real rephasing and non-rephasing data were added in order to obtain the 2D absorptive spectra reported in this work. No time domain filter was applied along $t_2$.

**Coherent 2DES Experimental Setup**

For the C-2DES experiments, a Ti:Sapphire regenerative amplifier (Spectra Physics Spitfire Pro, 1 kHz, 35 fs pulse duration) is used to pump a home-built degenerative optical parametric amplifier (DOPA) (12, 13), generating near-infrared pulses from 700-930 nm. The output of the DOPA is split with an 80/20 beamsplitter into the pump and probe, which are partially compressed with chirped mirrors (pump: -3000fs$^2$, 700-900nm, Femto Optics; probe: -280fs$^2$, 570-1040nm, Layertec). The pump is coupled into an acousto-optic pulse shaper (Dazzler, Fastlite), which is used to fully compress the pump to 11 fs and generate the pump pulse pair with controllable time delay ($t_1$) and relative carrier envelope phase ($\varphi_{12}$). The Dazzler also enables the time delay ($t_1$) to be scanned in the rotating frame by applying a time dependent carrier envelope phase. The probe is fully compressed to 11 fs with an SLM-based pulse shaper (Bio-photonics solutions FemtoJock P). To vary the waiting time ($t_2$), the pump pulse pair travels through a delay stage (Thor Labs DDS220). The pump and probe pulses are focused onto the sample in the pump-probe geometry using a curved mirror (f=250mm). At the sample position, the pump and probe spot sizes are 190 μm and 180 μm, respectively. The probe is coupled into a monochromator (Horiba iHR320, grating: 600 g/mm) and focused onto a CCD camera (Princeton Instruments Pixis 100). The pump and probe spectra, as well as the LH2 linear absorption spectrum are shown in Figure S8. The C-2DES experiments were collected in the annihilation-free regime, as confirmed by verifying that the transient absorption signal scaled linearly with the excitation energy. The pump and probe energy used were 7.3 nJ (at $t_1 = 0$) and 2.5 nJ, respectively. The time delay ($t_1$) was scanned in the partially rotating frame with a reference wavelength of 1000 nm from 0 to 180 fs in 5 fs steps. We also employed 0-π phase cycling of the pump pulses. To resolve the initial fast decay of the B850 diagonal peak (LDP), the waiting time ($t_2$) was scanned from -100 to 500 fs in 10 fs steps. To resolve the energy transfer from the B800 diagonal peak (UDP) to the B800-B850 cross peak (LCP), the waiting time ($t_2$) was scanned from -100 to 5000 fs in 100 fs steps. The C-2DES data was collected with all parallel polarizations. The same LH2 sample was used for the C-2DES and the F-2DES measurements. However, the concentration was increased for the C-2DES measurements such that the OD was 0.38 in a 0.2 mm pathlength flow cell. During the experiments, the sample was flowed with a peristaltic pump (Masterflex C/L) to prevent photobleaching.

After the nonlinear signal was isolated with 0-π phase cycling, the data was windowed along $t_1$ with a half tukey window (cosine fraction = 0.4). Then, the intensity at $t_1$=0fs was divided by 2 and the data was zero-padded to 256 points before Fourier transformation into the frequency domain. The frequency resolution of $\omega_1$ based on the sampling of $t_1$ was ~180 cm$^{-1}$ and the frequency resolution of $\omega_3$ was 1.36 cm$^{-1}$. To remove high frequency noise in the detection axis, the data was smoothed along $\omega_3$ with a 10 point moving mean.

**Absorption spectra before and after measurement for C-2DES and F-2DES**

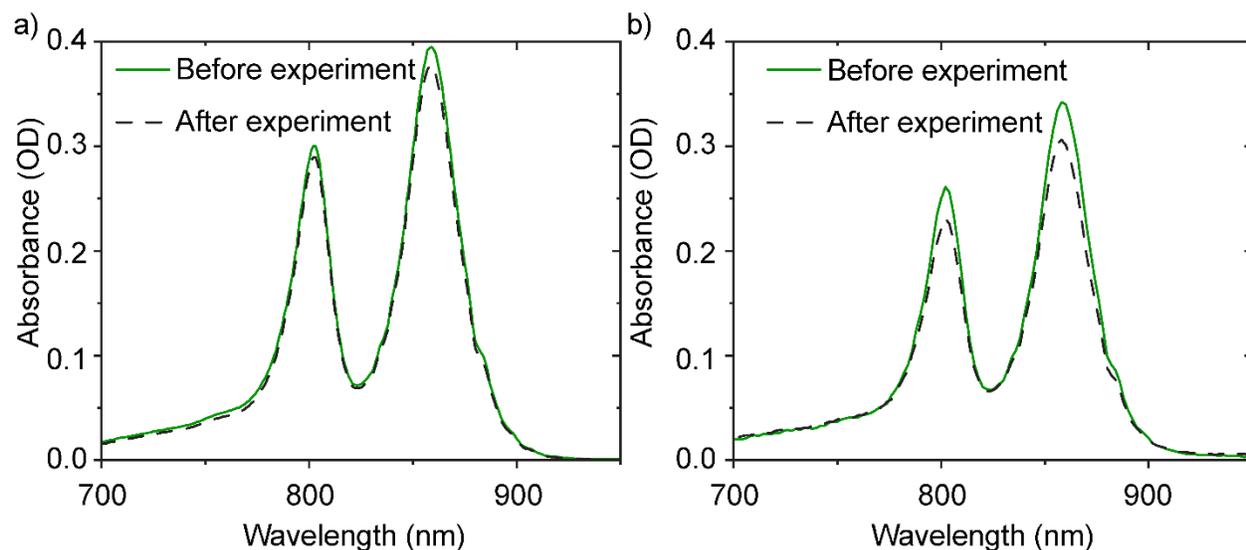

**Figure S7:** Absorption spectra of LH2 before and after the measurements for (a) coherently detected and (b) fluorescence-detected 2D experiments.

In Fig. S7a and Fig. S7b we show the absorption spectra before and after the coherently and fluorescence-detected 2D experiments, respectively. The complete experimental time was five days in the case of C2DES and three weeks in the case of F-2DES. In both experiments a slight amount of photobleaching is present which is more pronounced for the F-2DES experiments. The bleach leads to a reduction of the absorbance while the shape of the absorption spectrum stays constant. The timescale of one measurement is only a small fraction of this time, and therefore the photobleaching is not substantial enough to influence the reported dynamics and spectra.

## Experimental laser spectra

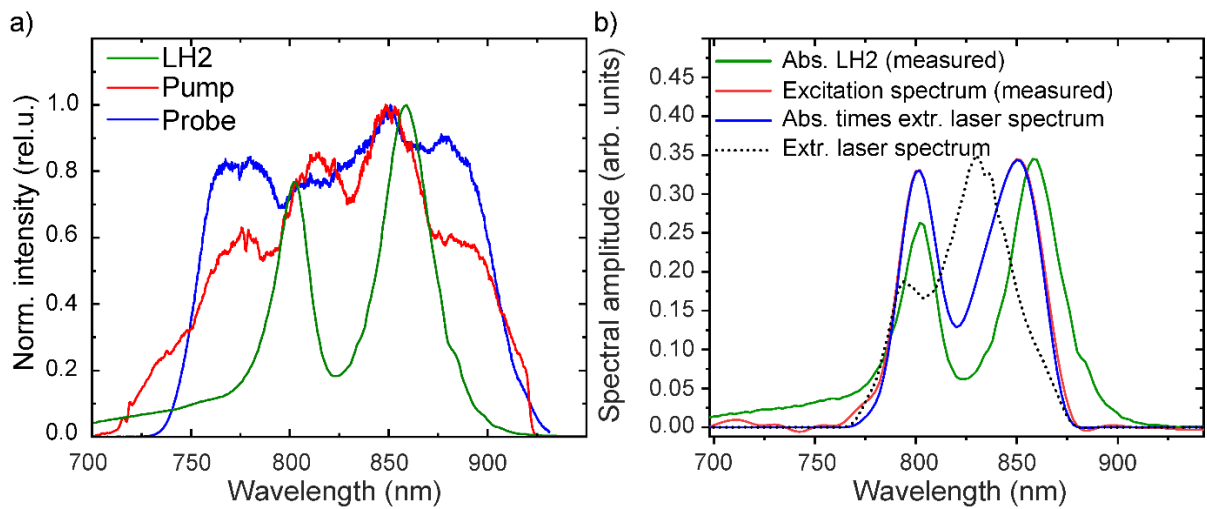

**Figure S8:** Absorption spectra of LH2 along with the laser spectra used in (a) C-2DES and (b) F-2DES.

In Fig. S8 we depict the laser spectra that were used in C-2DES (Fig. S8a) and in the case of F-2DES (Fig. S8b). In C-2DES the pump and the probe spectrum are generated from the same source, but are not identical since the pump beam contains an AOPDF pulse shaper to create a pulse pair with variable delay and the probe pulse is compressed with an SLM pulse shaper. We excite the sample with a pump spectrum with an FWHM with around 180 nm and probe the excited sample with a spectrum with similar width. Note that the probe spectrum has some spectral modulations. However, since we measure the transient change in absorbance the spectral shape of the probe spectrum does not influence our 2D spectrum. In the case of F-2DES, the pump and the probe spectra are almost identical since all four beams originate from the same white light source and travel through comparable optical elements. While the spectrum has a smaller width (~100 nm) compared to the spectrum used in C-2DES, we still cover the main transitions of LH2. The laser spectrum was isolated by comparing the two-pulse fluorescence excitation spectrum (red in Fig. S8b) acquired in the same measurement as the F-2DES with the measured absorption spectrum (green in Fig. S8b). The B800 peak is nicely covered by the laser spectrum, whereas the red edge of the B850 transition is not excited by the laser. The red edge was purposefully chosen to excite most of the transition while also allowing for spectral separation of the laser from the fluorescence of LH2. The extracted laser spectrum (black dotted in Fig. S8b) was used for the theoretical calculations.

## Coherent 2DES Dynamics

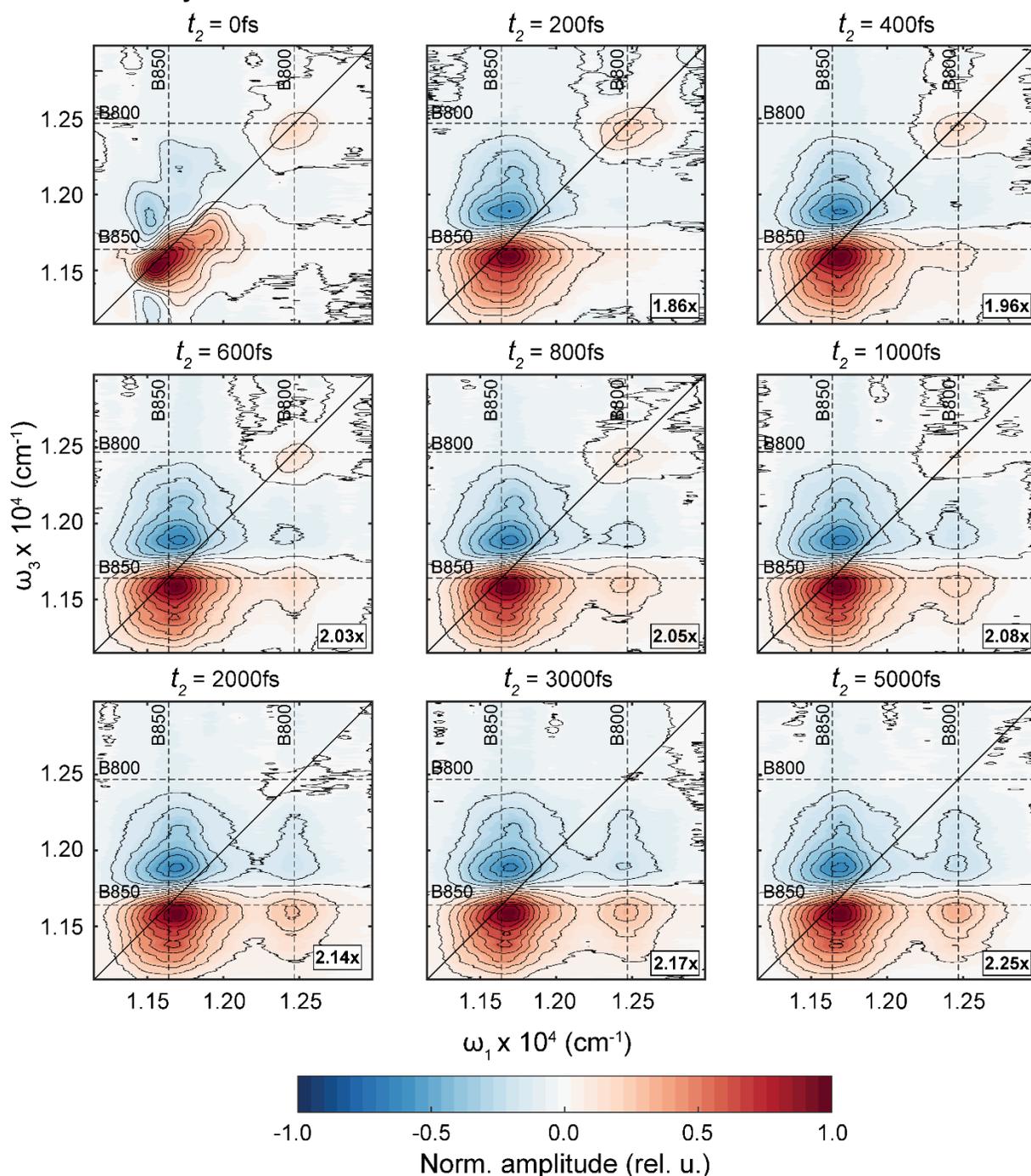

**Figure S9:** Coherent two-dimensional spectra of LH2 for several waiting times $t_2$. Each spectrum is scaled by the factor given in the bottom right corner of each individual spectrum and normalized to 1.

In Fig. S9, C-2DES spectra of LH2 are shown for several waiting times. The appearance of the C-2DES spectra is distinct from the F-2DES spectra since positive GSB and SE, as well as negative ESA features are visible in C-2DES. The lower diagonal peak corresponding to B850 band is dominating since the absorption of this band is higher than the B800 band. Furthermore, our pump spectrum is more intense at B850 increasing the expected intensity difference in the 2D spectrum between the B800 and B850 diagonal peaks. For later waiting times a clear decay of the B800 upper diagonal peak is visible and the lower diagonal cross peak rises.

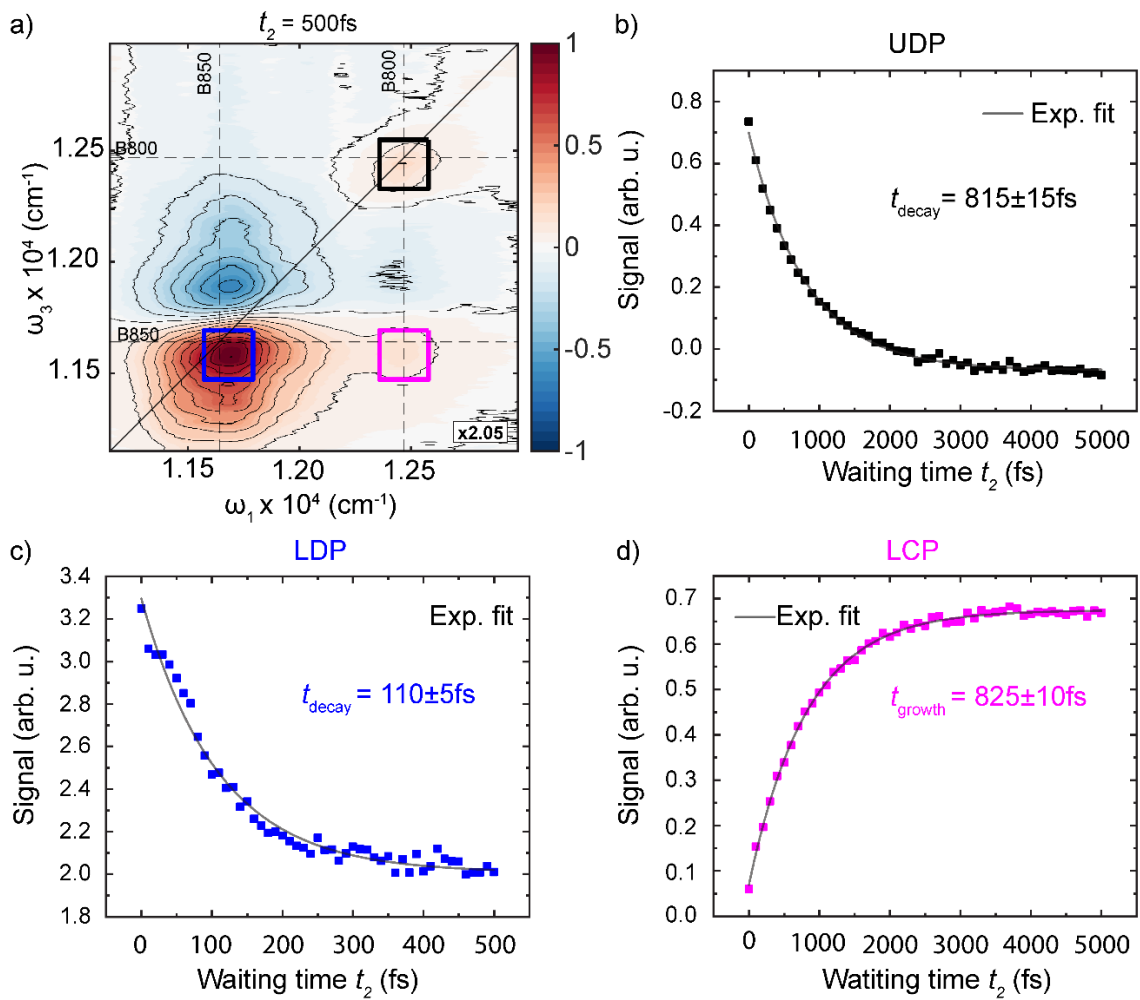

**Figure S10:** Coherent 2DES of LH2. a) C-2DES spectrum at a waiting time of 500 fs. The squares mark the areas for averaging to obtain the traces shown in b)-d). b) Decay and exponential fit (black solid) of the upper diagonal peak (B800-B800). c) Exponential decay of the lower diagonal peak (B850-B850) and exponential fit (black solid line). d) Exponential rise and exponential fit (black solid line) of the cross peak resulting from exciting at B800 and probing at the corresponding wavelength of B850.

In Fig. S10a the marked squares indicate the regions over which signal was averaged to obtain the waiting time $t_2$-dependent traces shown in Fig. S10b-d. The UDP (Fig. S10b) and the LCP (Fig. S10d) decay or rise with almost identical time constants of 815 and 825 fs, respectively. Interestingly, the signal of the UDP is slightly negative at the end due to the extended ESA of the LCP that is reaching into the region of interest of the UDP. To resolve the ultrafast relaxation within the B800 diagonal peak. We performed additional measurements with smaller step size in $t_2$ compared to the measurements shown in Fig S10b and d. We resolve an exponential decay of the B800 diagonal peak of 110 fs that we attribute to ultrafast intraband relaxation. We also noticed coherent oscillations on the signal with a small amplitude. However, a detailed analysis of these features is beyond the scope of this work.

## Calculation of Absorption and F-2DES Spectra

The Hamiltonian in diabatic (site) basis is described by the site energies $e_i$ and electronic coupling between the transitions $J_{nm}$:

$$H = \sum_{n,m} |e_n\rangle (e_n \delta_{nm} + J_{nm}) \langle e_m|. \qquad (S7)$$

The coupling is calculated based on the crystal structure and considered constant. The site energies are taken to be statically disordered with a normal distribution around their mean values. For each realization of the energetic disorder, the excitonic Hamiltonian $H_{nm}$ is diagonalized:

$$H_{nm} = \sum_{i,j} c_n^i H_{ij}^{ex} c_m^j, \qquad (S8)$$

Where $H^{ex}$ is diagonal with the exciton state energies $\epsilon_i$ as eigenvalues, and $C_{ni}$ are the elements of the transformation matrix with *n* the site index and *i* excitonic state index. The transition dipole moments (whose sizes are the oscillator strengths of the respective excitonic transitions) are transformed into the exciton basis as

$$\mu_i = \sum_n |c_n^i| \mu_n. \qquad (S9)$$

The values of the electronic coupling $J_{nm}$ determine the diagonalization $c_n^i$ and thus the excitonic state structure as determined by the excitonic energies $\epsilon_i$, oscillator strengths $|\mu_i|^2$ and the excitonic delocalization (determined by $|c_n^i|^2$).

The vibrational environment is described as a bath with a spectral density $C(\omega)$, taken as a sum of three overdamped Brownian oscillators (14):

$$C(\omega) = \sum_{s=1}^{3} \left(1 + \cotanh \frac{\hbar \omega}{2 k_B T}\right) \frac{2 \lambda_s \Lambda_s \omega}{\omega^2 + \Lambda_s^2}. \qquad (S10)$$

Here, $\lambda_s$ and $\Lambda_s$ are the bath reorganization energies and inverse correlation times, respectively, with values taken the same as in (14). The spectral density is used for calculation of the Redfield population transfer rates from state *j* to state *i*:

$$k_{ij} = \sum_n |c_n^i|^2 |c_n^j|^2 C(\omega_{ji}). \qquad (S11)$$

The spectral density is also used to evaluate the lineshape function in a high-temperature limit (15)

$$g(t) = \sum_{s=1}^{3} \frac{2 \lambda_s k_B T}{\hbar \Lambda_s^2} [\exp(-\Lambda_s t) + \Lambda_s t - 1]. \qquad (S12)$$

The lineshape of the *i-th* exciton is given by this lineshape including exchange narrowing, and relaxation-induced dephasing:

$$L_i(t) = \exp(-i \omega_i t - g_i(t) - \Gamma_i t), \qquad (S13)$$

where $g_i(t) = \sum_n |c_n^i|^4 g(t)$ and $\Gamma_i = -\frac{1}{2} k_{ii} = \frac{1}{2} \sum_{i \neq j} k_{ji}$ is all the rates away from the state *i*. Since the relaxation-induced dephasing leads to a Lorentzian lineshape, it makes sense to approximate the lineshape itself by a Lorentzian as well, see Fig. S11a) in the Parameters section. The excitonic coherence dephasing is calculated as

$$\gamma_{ij} = \sum_n |c_n^i|^2 |c_n^j|^2 C(0) - \frac{1}{2}(k_{ii} + k_{jj}). \qquad (S14)$$

Here, the first term is the pure dephasing and the second one the relaxation-induced dephasing.

The absorption spectrum is calculated as

$$\alpha(\omega) = \sum_i |\mu_i|^2 \omega L_i(\omega).\qquad (S15)$$

The 2D spectrum is calculated directly in the frequency domain using pre-calculated lineshapes and orientationally-averaged dipole-moment combinations, summing up the response pathways. We thus have

$$\mathrm{F-2DES}(\omega_1, t_2, \omega_3) = \sum_{ij} \left[ |\mu_i|^2 |\mu_j|^2 + 2(\mu_i \cdot \mu_j)^2 \right] \left[ 1 + U_{jj,ii}(t_2) + \exp(-\gamma_{ij} t_2) \cos(\omega_{ij} t_2) \right] L_i(\omega_1) L_j(\omega_3).$$
$$\mathrm{S}(16)$$

Here, the population transfer propagator $U_{jj,ii}(t_2)$ is calculated by the matrix exponential using the rate matrix. We consider the same transfer rate $k_T$ from all states of manifold 1 (i.e., B800) to manifold 2 (i.e., B850):

$$k_{i \in B850, j \in B800} = \frac{1}{N_{B850}} k_T \, \forall i, j. \qquad (S17)$$

*Parameters for calculations:*

We have used the same excitonic model as in our previous work on LH2 which simulated linear, nonlinear and single-molecule spectra (14). That work built upon the preceding models (16, 17). For a detailed description, we refer the reader to these publications. Briefly, we calculated the electronic transition coupling in a dipole-dipole approximation based on the LH2 crystal structure. The nearest-neighbour coupling was 293 cm$^{-1}$. Site energies were taken to be (12230 cm$^{-1}$ and 12430 cm$^{-1}$) for B850 dimers and 12520 cm$^{-1}$ for B800 sites. These are approximately 100 cm$^{-1}$ redshifted from our previous work since we do not consider Stokes shift in this work for simplicity. Site energy disorder was taken to be a 300 cm$^{-1}$ wide Gaussian distribution (width sigma) in B850 ring, and 0.31 times that (93 cm$^{-1}$) in B800. This agrees with the observation by Cupellini et al. that the B850 coupling has charge transfer character and is thus more sensitive to fluctuations (18). These parameters already determine the excitonic state structure and thus the expected SE/GSB ratio in spectrally integrated manifolds.

Both the lineshape and dynamics are given by the interaction with the vibrational bath, which we describe by a three-component spectral density as in ref. (14), with parameters $\lambda_s = \{15, 135, 165\}$ cm$^{-1}$ and $\Gamma_s = \{30, 400, 1200\}$ cm$^{-1}$. The population transfer rates as well as inter-exciton coherence pure dephasing are described by Redfield theory as formulated above. The transfer from B800 to B850 rings is taken to be site-non-specific with a 830 fs time constant, determined from the C-2DES measurement. The excitonic lineshape is calculated by cumulant expansion and fitted with a Lorentzian (see Fig. S11) for faster calculation in the spectral domain, the fitted width in the site basis is $\Gamma = 185$ cm$^{-1}$. The excitonic lineshapes take into account exchange narrowing by exciton delocalization, as well as relaxation-induced dephasing. The spectra are calculated as described above, and further multiplied by the laser spectrum. The 2D spectrum was finally convoluted with the Fourier transform of the time-filter window function described in the data processing section above. The window function in the time domain is $w(\tau) = \frac{1}{2}\left\{1 - \tanh\left[\frac{\tau - \tau_0}{\sigma}\right]\right\}$, where $\tau_0 = 84$ fs and $\sigma = 7$ fs.

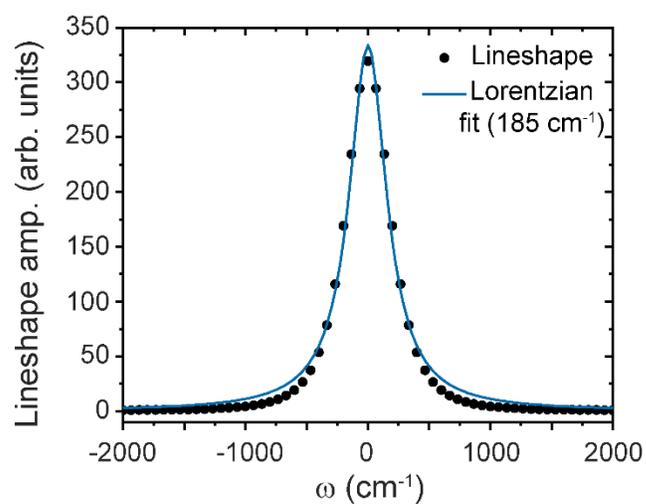

**Figure S11:** Spectral properties for calculations. Site lineshape calculated by a cumulant expansion (black dots) fitted by a Lorentzian (blue line) that was used for calculating the spectra.

**Initial Amplitude in Fluorescence-Detected 2DES**

A striking feature of the dynamics in the F-2DES spectra is the high signal amplitude at $t_2 = 0$. At this particular time point the time ordering of the pulses breaks down, together with other effects such as cross-phase modulation or two-photon absorption (19-21). In the literature these effects are summarized in the term "coherent artifact" (15). However, despite all the different effects, we can clearly see in Fig. S10 c-d that the initial spike is absent in the C-2DES measurement. Therefore, an additional effect has to take place in F-2DES spectroscopy that is absent in its coherently detected counterpart. We identified one of these effects as coherences at $t_2 = 0$ as depicted in the representative double-sided Feynman diagrams in Fig. S12. The important point is that pathways with coherences will contribute even if the dephasing time is extremely short, for example, if the states are spatially separated. These pathways can be viewed as the product of two separate pathways. For example, in the case of the lower diagonal peak (Fig. S12, lower left), one pathway in which pulse 1 and 3 excite and de-excite a coherence $|0\rangle\langle B850_j|$ and a second pathway in which pulse 2 and 4 create a population $|B850_l\rangle\langle B850_l|$. Note that there also exist two additional ESA pathways with the same coherence at $t_2$. However, in the case of efficient annihilation these two pathways cancel each other.

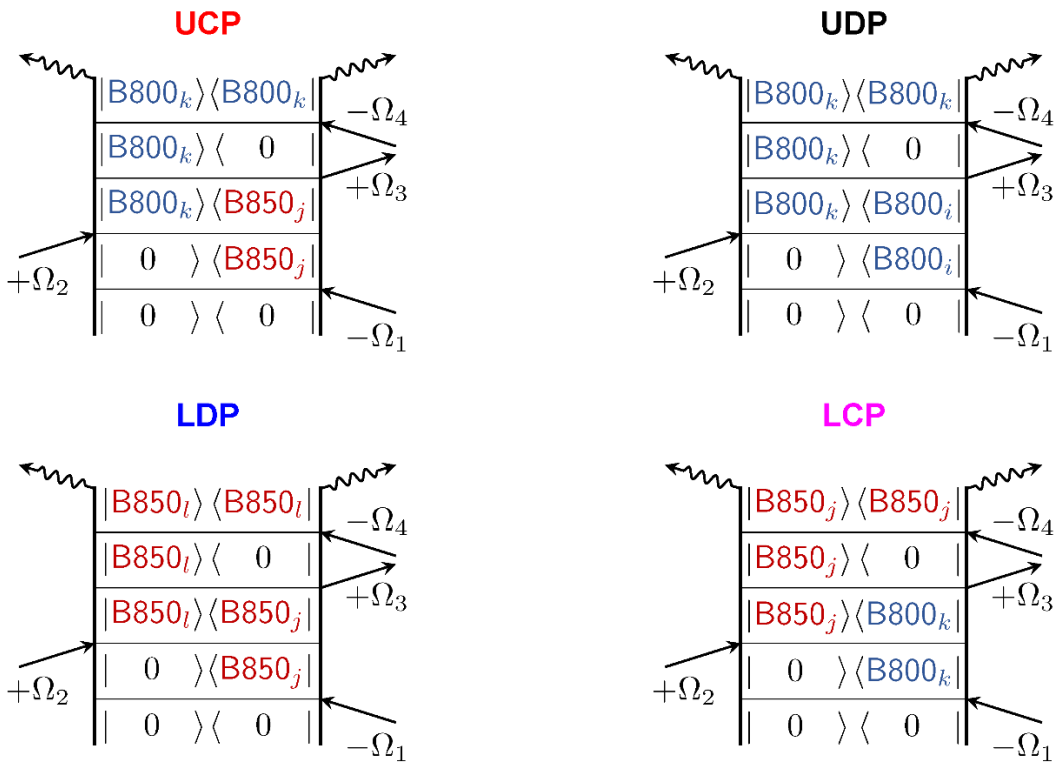

**Figure S12:** Examples of double-sided Feynman diagrams contributing at $t_2 = 0$ to the four peaks in LH2.

We calculated the F-2DES spectra with and without these coherences to demonstrate that the initial spike in the F-2DES signal which rapidly decays is not caused by pulse overlap, but by the inter-excitonic coherence term. The short laser pulses create superpositions of the excited excitonic states (*e.g.*, *i* and *j*), which oscillate at the difference frequencies $\omega_{ij}$. This superposition decays with the decoherence rate $\gamma_{ij}$ given above. On top of that, averaging over the measured ensemble of LH2 complexes leads to a distribution of oscillation frequencies (due to the static disorder) and thus to a rapid dephasing. The result is a rapidly decaying contribution at all the peak positions. This is demonstrated on LH2, where leaving the coherence terms out of the calculation results in the absence of the initial spike (Fig. S13).

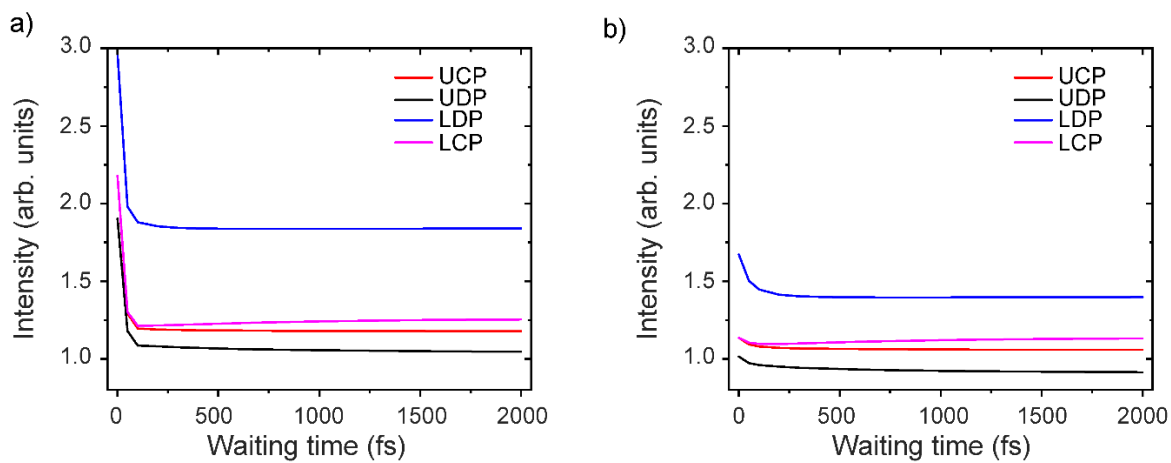

**Figure S13:** Comparison of peak intensity traces for LH2 calculated with (a) and without (b) inter-excitonic coherence. Without the inter-excitonic coherence, the initial spike in the data around $t_2 = 0$ is absent.

**Effect of laser spectrum on final F-2DES spectra**

In the experimentally measured F-2DES spectra, the UDP falls nicely along the diagonal at the expected peak from the linear absorption measurement as indicated by the dashed lines in Fig. 1 of the main text. However, the LDP is not centered at the expected position from the linear absorption, with the observed peak position blue shifted. This is a result of the laser spectrum not fully covering the B850 transition, as seen in Fig. S8b). To further understand the role of the laser spectrum in F-2DES, we simulated the F-2DES spectra as a function of waiting time with lineshapes modifed by the laser spectrum and the real lineshapes from the excitonic coupling in LH2. Fig. S14 shows the simulated F-2DES spectra at three waiting times with and without modification by the experimental laser spectrum. Both diagonal peaks in the 2D spectra with the real lineshapes are observed at the expected positions from the linear absorption spectrum of LH2, given by the intersection of the dashed lines. However, upon modulating the F-2DES spectra by the laser spectrum, the LDP blue shifts and there is more intensity between the peaks such that the intensity corresponding to specific transitions (B800 vs B850) is more difficult to separate in frequency space.

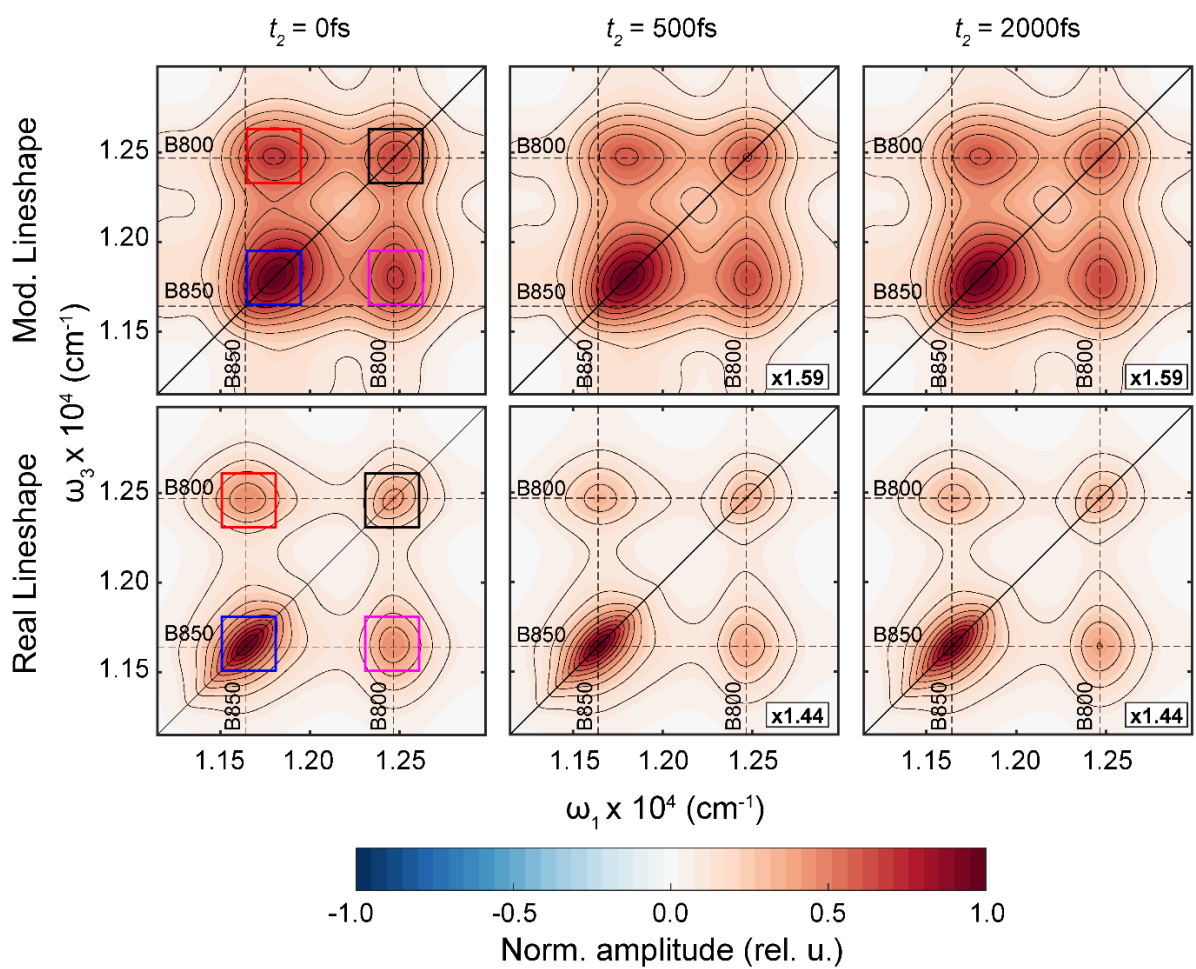

**Figure S14:** Simulated F-2DES spectra with (a) the lineshape modified by the experimental laser spectrum and (b) the real lineshapes from the excitonic structure. The squares indicate the regions integrated to extract the waiting time dependence of the four peaks.

With spectrally overlapping excitonic states, it can be difficult if not impossible to distinguish the SE from GSB, preventing us from observing the excited state dynamics even when its relative contribution is substantial. Fortunately, in many large aggregates such as photosynthetic complexes, there exist several sub-groups of pigments which can be distinguished spectrally. These are often strongly-coupled with fast inter-manifold population relaxation, while the dynamics between the segments can be, due to their weak coupling and energy separation, significantly slower. In this case, a separation of timescales applies, and we can discuss the SE/GSB of the quasi-equilibrated individual manifolds. The SE can in this case be clearly distinguished as the slowly decaying component at the position of the higher-energy segment, and as a slowly rising component at the cross-peak position between the segment spectra. LH2 with its two well-separated rings is a perfect example. Considering the separation of timescales for intra- and inter-ring relaxation, the equilibrated equation from the preceding section becomes augmented by the population transfer. For the B800 state decay visible in the UDP, we obtain

$$\left(\frac{\text{SE}}{\text{GSB}}\right)_{\text{UDP}} = \frac{\sum_{i,k=1}^{9}|\mu_{B800_i}|^2 P_k^{eq}|\mu_{B800_k}|^2}{\sum_{i,k=1}^{9}|\mu_{B800_i}|^2|\mu_{B800_k}|^2} e^{-k_{B800\to B850}t_2} = \frac{\sum_{k=1}^{9} P_k^{eq}|\mu_{B800_k}|^2}{\sum_{k=1}^{9}|\mu_{B800_k}|^2} e^{-k_{B800\to B850}t_2}. \tag{S18}$$

For the B850 population rise at the LCP position, we get

$$\left(\frac{\text{SE}}{\text{GSB}}\right)_{\text{LCP}} = \frac{\sum_{j=1}^{18} P_j^{eq}|\mu_{B850_j}|^2}{\sum_{j=1}^{18}|\mu_{B850_j}|^2}(1 - e^{-k_{B800\to B850}t_2}). \tag{S19}$$

Note, that since the transferring population of the B800 ring does not depend on the initial state, the $|\mu_{B800_i}|^2$ factors out of the fraction. In a similar way, SE can be identified in many photosynthetic complexes, and the inferred SE/GSB ratio can be used to learn about the excitonic state structure and delocalization.

## Comparison of SE to GSB Ratio

As discussed in the main text, only the SE contains information about the energy transfer dynamics and the GSB is a constant background. Fig. S14 shows that the laser spectrum significantly changes the final F-2DES spectra. In Fig. S15 we show the corresponding $t_2$-dependent traces extracted by integrating over the boxed regions of the spectra shown in Fig. S14, along with the experimental data. The UCP and LDP peak amplitudes were normalized such that the average amplitude for long waiting times ($t_2$ >1ps) was 1. The UDP and LCP were normalized based on the exponential fits, which are described below. The dynamics of the UCP and LDP are not significantly affected since they do not report on energy transfer as seen by only small differences in the experiment and the theory with and without the laser spectrum in Fig. S15a and c. However, the distortion of the spectrum by the laser spectrum results in the observed kinetics of the UDP and LCP being fainter than expected given the excitonic structure of LH2, which is illustrated by the real lineshape traces deviating from the experiment values and the modfed lineshape traces in Fig. S15b and d. To precisely determine the percent rise/decay of the LCP/UDP, we fit the data to extract the amplitude of the SE and GSB signals.

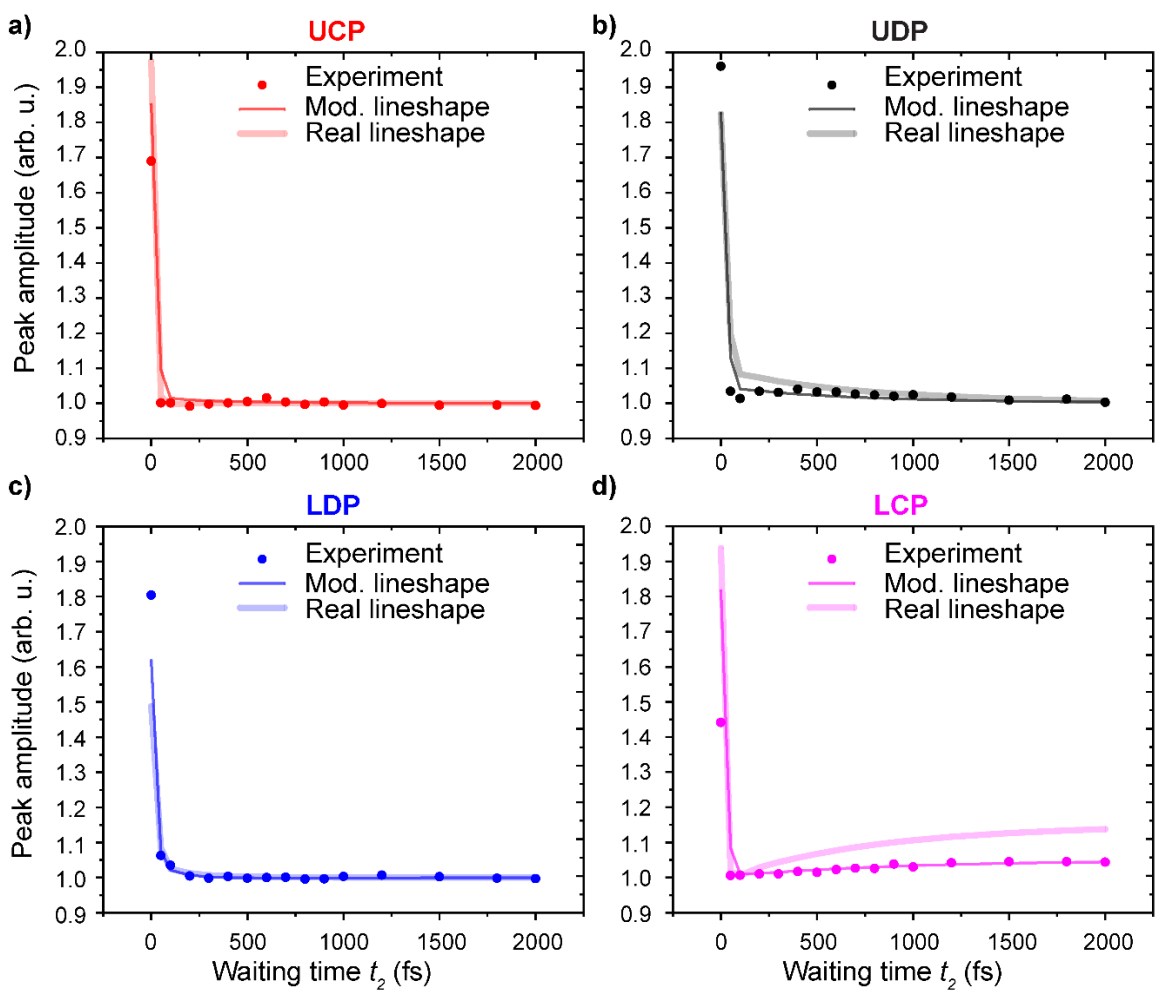

**Figure S15:** Waiting time $t_2$-dependent traces for the experimental F-2DES data (dots), simulated F-2DES data with modified lineshapes by the laser spectrum (thin, dark line) and simulated F-2DES data with the real lineshapes (thick, light line) for the (a) UCP, (b) UDP, (c) LDP, and (d) LCP.

To achieve a high enough signal to noise ratio in the experimental F-2DES data to extract the energy transfer timescales and SE to GSB ratio, three different data sets needed to be averaged. The averaged data along with the standard deviation from the multiple measurements is the experimental data shown in

Fig. 2 in the main text. Prior to averaging the three measurements together, the absolute intensity for each measurement and peak was normalized such that the average intensity between 50 fs and 2000 fs was one to account for slight differences in signal level between measurements. The $t_2$ dependent intensity of the four peaks for each of the three scans and the mean are shown in Fig. S16. To extract the ratio of the SE to GSB for the LCP and UDP from the experimental and simulated F-2DES data, we fit the data to an exponential rise and decay respectively:

$$I_{\text{LCP}}(t_2) = A_{\text{SE}}\left(1 - e^{-\frac{t_2}{\tau}}\right) + A_{\text{GSB}} \qquad (S20)$$

$$I_{\text{UDP}}(t_2) = A_{\text{SE}} \cdot e^{-\frac{t_2}{\tau}} + A_{\text{GSB}} \qquad (S21)$$

where $A_{\text{SE}}$ and $A_{\text{GSB}}$ are the amplitudes of the stimulated emission and ground state bleach and $\tau$ is the energy transfer time. The standard deviation of the three measurements was used to weight the experimental data for the fits. The results from the fits were also used to normalize the data in Fig. 2 of the main text and Fig. S15. The peak amplitude for both the LCP and UDP were divided by $A_{\text{GSB}}$. Thus, the normalized peak amplitudes are given by

$$I_{\text{LCP}}^{\text{norm}}(t_2) = \frac{I_{LCP}(t_2)}{A_{GSB}} = \frac{A_{SE}}{A_{GSB}}\left(1 - e^{-\frac{t_2}{\tau}}\right) + 1 \qquad (S22)$$

$$I_{\text{UDP}}^{\text{norm}}(t_2) = \frac{I_{UDP}(t_2)}{A_{GSB}} = \frac{A_{SE}}{A_{GSB}} \cdot e^{-\frac{t_2}{\tau}} + 1 \qquad (S23)$$

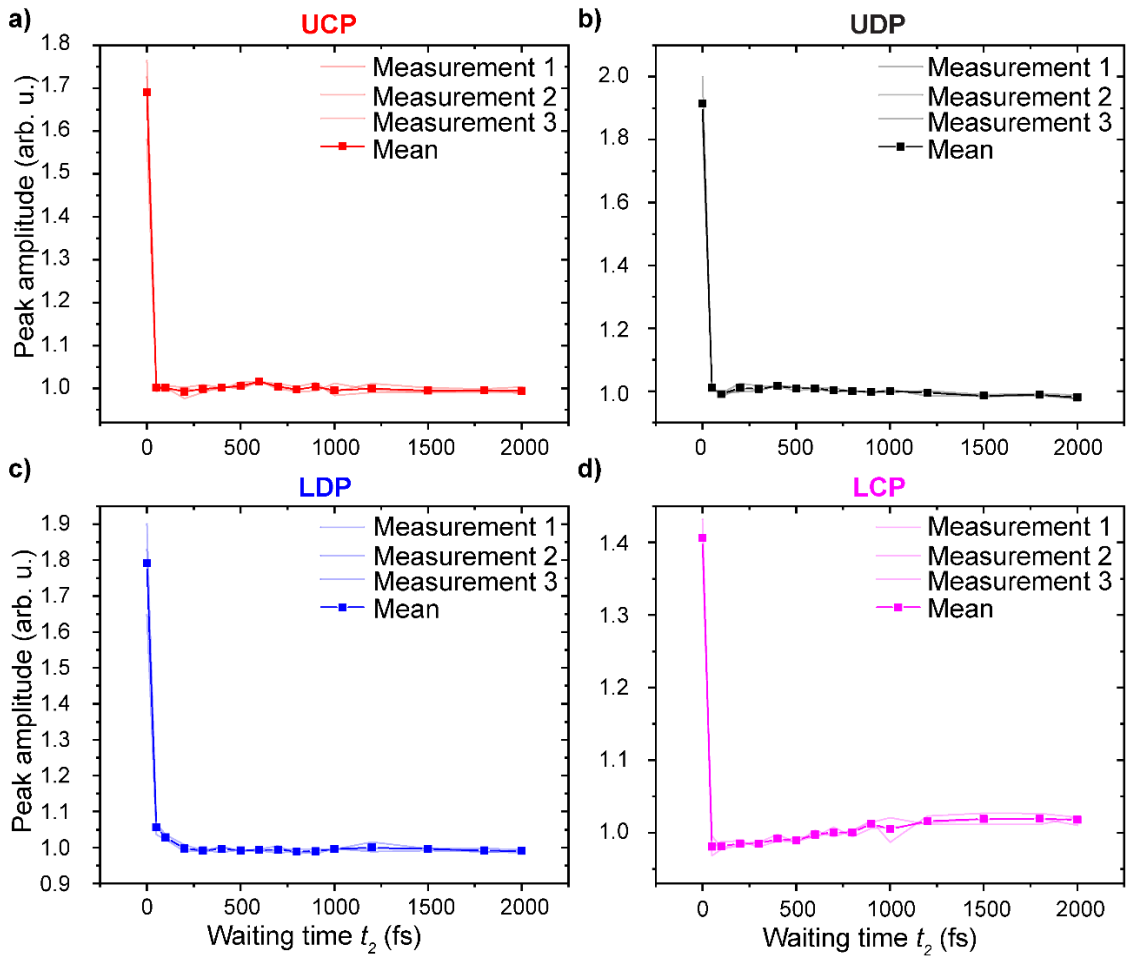

**Figure S16:** Waiting time $t_2$-dependent traces for the three individual F-2DES measurements and the average for the (a) UCP, (b) UDP, (c) LDP, and (d) LCP.

From the values of A_SE and A_GSB from the fits, we calculated the percent rise/decay by $\frac{A_{SE}}{A_{GSB}} \cdot 100$. An overview of the results from the experimental fit and theory using the excitonic model (with modification by the laser spectrum and with the real lineshape) and combinatoric treatment (1/*N* limit) are shown in Table S1 below. While the experimental values for the LCP agrees well with the 1/*N* limit, the UDP result is very far off indicating that LH2 cannot be described with the combinatoric argument for *N* molecules (6). When considering the LCP and UDP changes together, we have much better agreement using an excitonic model for LH2 where the spectral lineshapes are modified by the laser spectrum. Without the spectral modification due to the laser, the growth of the LCP would be ~3x larger than the 1/*N* limit, while the decay of the UDP would be slightly lower.

|  | LCP | UDP |
|---|---|---|
| **Experiment** | 6.2% | 4.5% |
| **Experiment: Fixed time constant to 830fs** | 5.3% | 3.7% |
| **Theory: Excitonic Model – Modifed Lineshape** | 4.8% | 4.1% |
| **Theory: Excitonic Model – Real Lineshape** | 15.2% | 8.5% |
| **Theory: 1/*N* limit** | 5.6% | 11.1% |

**Table S1:** Comparison of SE to GSB ratio from fits to the experimental data versus the excitonic model and combinatoric 1/N theory results for the LCP and UDP.

**Effect of Polarization on SE/GSB ratio**

Considering response pathways with population dynamics in $t_2$ only, together with efficient EEA so that excited-state absorption does not lead to increased emission and is accordingly absent from the spectra, one gets for the SE/GSB ratio

$$\frac{\text{SE}}{\text{GSB}} = \frac{\sum_{i,j=1}^{N} \langle \mu_i \mu_i \mu_j \mu_j \rangle U_{ji}(t_2)}{\sum_{i,j=1}^{N} \langle \mu_i \mu_i \mu_j \mu_j \rangle}. \qquad (S24)$$

Here, $U_{ji}(t_2)$ is the waiting-time population propagator (conditional probability that excitation of state $i$ ends up in state $j$ after time $t_2$), $\mu_i$ are the transition dipole moment vectors, and $\langle \rangle$ denotes orientational averaging determined by the pulse polarization geometry. For magic angle, we have $\langle \mu_i \mu_i \mu_j \mu_j \rangle_{ma} = \frac{1}{9} |\mu_i|^2 |\mu_j|^2$, while for all-parallel we have $\langle \mu_i \mu_i \mu_j \mu_j \rangle_\| = \frac{1}{15} \left[ |\mu_i|^2 |\mu_j|^2 + 2(\mu_i \cdot \mu_j)^2 \right]$.

Now let us look at several limits for which the expression simplifies. If the final population of all states is the same regardless of the initial state, we have $U_{ji}(t_2) = \frac{1}{N} \forall i,j$, which can be taken out of the sum. The ratio is then, regardless of the polarization,

$$\left( \frac{\text{SE}}{\text{GSB}} \right)_{\text{uniform population}} = \frac{1}{N}. \qquad (S25)$$

If all dipole moments are the same, the averaged dipoles can be taken out of the sum, and because $\sum_{i,j=1}^{N} U_{ji}(t_2) = N$ for all propagators and $\sum_{i,j=1}^{N} 1 = N^2$, one also gets

$$\left( \frac{\text{SE}}{\text{GSB}} \right)_{\text{same dipoles}} = \frac{1}{N}. \qquad (S26)$$

Under the magic angle condition, both the dipole magnitudes and their populations thus have to be different in order for the SE/GSB ratio to differ from the 1/$N$ limit, regardless of the dipole orientation. In the all-parallel setting, there is an additional anisotropy contribution which can slightly alter the ratio as well, dependent on the dipole orientation.

The interesting case relevant to our situation is that of (thermally) quasi-equilibrated excitonic manifold, in which case $U_{ji}(t_2) = P_j^{eq} \forall i$ with the additional possibility of slow dynamics in $t_2$ (*e.g.*, between the B800 and 850 rings, as we discuss below). The equilibrium populations at temperature T are $P_j^{eq} = \exp\left( -\epsilon_j / k_B T \right) \left( \sum_{j=1}^{N} \exp\left( -\epsilon_j / k_B T \right) \right)^{-1}$, where $\epsilon_j$ are the state energies. For the magic angle, one then has

$$\left( \frac{\text{SE}}{\text{GSB}} \right)_{ma} = \frac{\sum_{j=1}^{N} |\mu_j|^2 P_j^{eq}}{\sum_{j=1}^{N} |\mu_j|^2}, \qquad (S27)$$

since the sums can be factorized and the sum $\sum_{i=1}^{N} |\mu_i|^2$ factors out. In the all-parallel experiment, this is not the case:

$$\left( \frac{\text{SE}}{\text{GSB}} \right)_\| = \frac{\sum_{i,j=1}^{N} \left[ |\mu_i|^2 |\mu_j|^2 + 2(\mu_i \cdot \mu_j)^2 \right] P_j^{eq}}{\sum_{i,j=1}^{N} \left\{ |\mu_i|^2 |\mu_j|^2 + 2(\mu_i \cdot \mu_j)^2 \right\}}. \qquad (S28)$$

This expression cannot be simplified in general, only for special cases. One can, however, always calculate the SE/GSB ratios for both magic angle and parallel and compare. For all-parallel dipoles (both in J- and H-aggregate), there is no difference. In case of symmetric aggregates, the difference is expected to be minimal, since the signal can be viewed as the isotropic one plus a small anisotropy, which by itself has a similar SE/GSB ratio. This is shown for the example of a circular aggregate (tangential dipole orientation, $N = 18$) below in Fig. S17. Parameters of the aggregate are the same as used for the circular aggregate in Fig. 4A of the main text.

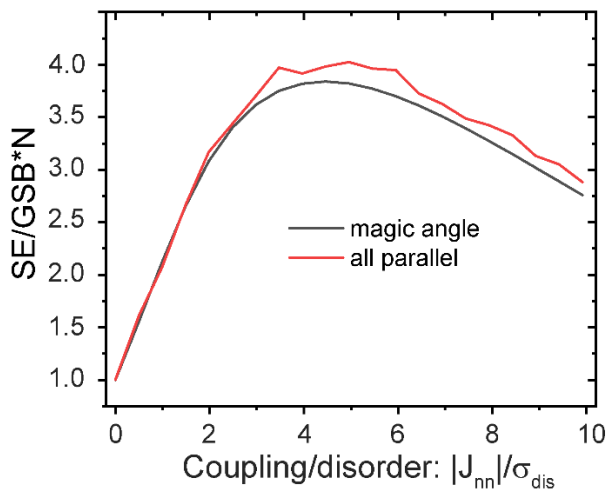

**Figure S17:** SE/GSB ratio dependence on electronic coupling vs disorder, comparison of isotropic vs anisotropic signal. The data are calculated for a circular $N = 18$ aggregate. Clearly, the anisotropy increases the SE/GSB ratio only marginally.

## Simulations of Coupled Aggregates

The expressions for the SE/GSB ratio derived in the sections above are very simple and depend on the static properties of the excitonic manifolds only, without any dynamics or line shapes. In order to verify their applicability to realistic systems, we calculated the F-2DES spectra of LH2 and other molecular aggregates. For LH2 the results are shown in the main text, and the quantitative comparison with the expressions (Eq. (4),(5) of the main text) was complicated by the finite laser bandwidth effects and overlapping states from the two rings. To better understand the role of spectral separation versus excitonic coupling on the appearance of dynamics, and demonstrate that the extracted SE/GSB ratio can be indeed inferred from the excitonic structure, we calculate model aggregates with two well-separated excitonic manifolds and varying geometry. The donor molecules are weakly coupled with all states similarly bright, whereas the acceptor molecules are more strongly coupled with one much brighter state, as depicted by the oscillator strengths in Fig. S18a).

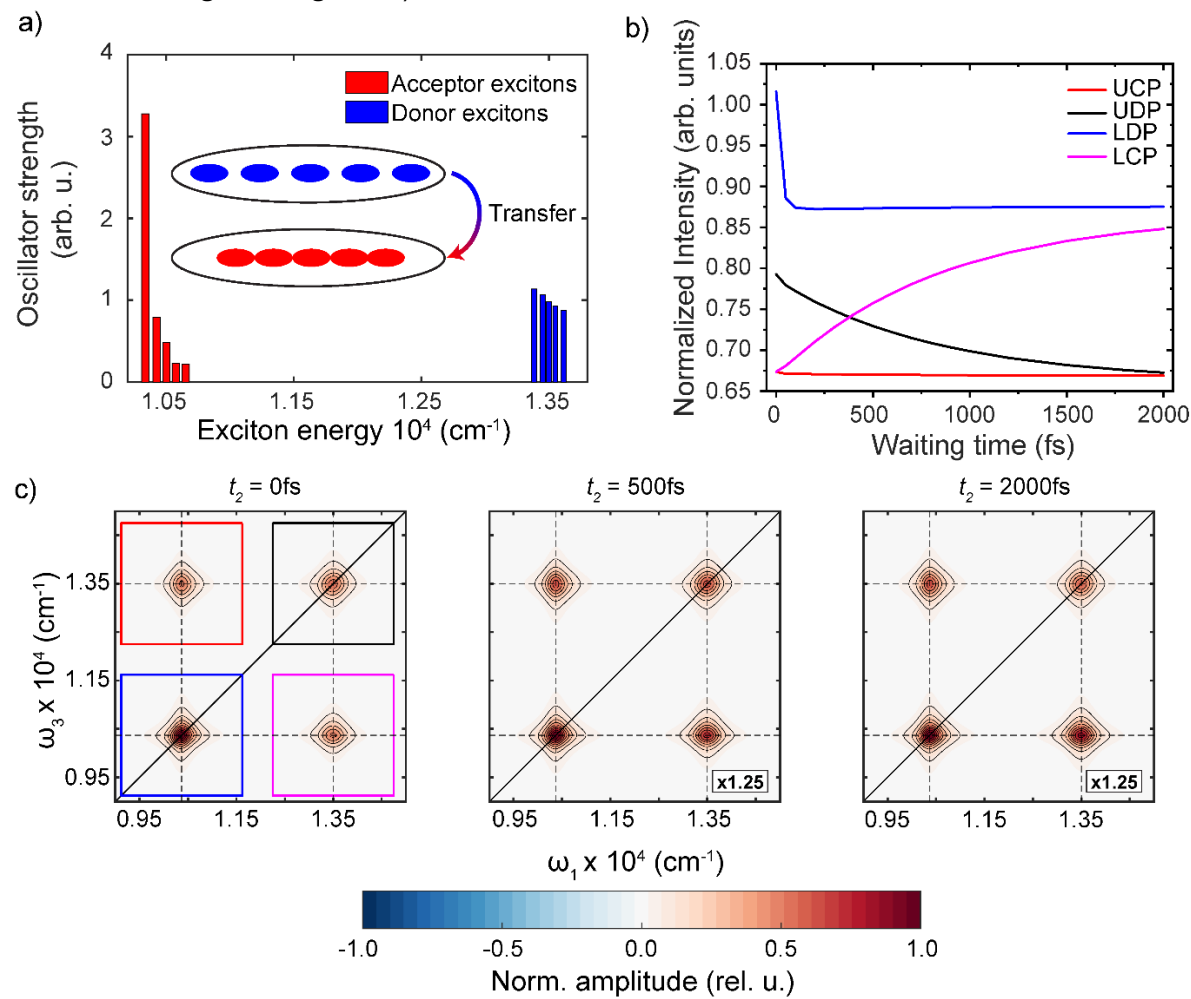

**Figure S18:** Model system of two coupled aggregates (composed of 5 molecules each) that undergo energy transfer. (a) The oscillator strength of the excitons for the donor and acceptor aggregates, (b) time traces of each of the four peaks, and (c) simulated F-2DES spectra.

For the calculation of F-2DES spectra, we use the same model and parameters for the LH2, with these differences: the site energies are taken to be 10500 cm$^{-1}$ and 13500 cm$^{-1}$ in the two manifolds. The energetic disorder is taken to be 100 cm$^{-1}$. We consider a circular geometry (tangential dipole orientation), as well as linear H-type and J-type in each manifold, same as in Fig. 4a in the main text. For all the aggregates we consider a 830 fs transfer between the manifolds. We have simulated the F-2DES spectra of the aggregates and verified that the observed SE/GSB ratios inferred from the UDP decay and LCP rise agree with the prediction of Eq. 4 and 5 in the main text, respectively. In Fig. S18 one such exemplary J-type aggregate is shown for $N = 5$ sites in both manifolds. The higher-energy manifold features a larger inter-pigment distance (weaker coupling $\frac{|J_{nn}|}{\sigma_{dis}} = 0.02$) and the lower one with pigments closer together (stronger coupling $\frac{|J_{nn}|}{\sigma_{dis}} = 0.57$). Clearly, in the strongly coupled manifold the oscillator strength is redistributed into the lowest-energy state (J-type behavior), while in the weakly-coupled manifold all states are more or less the same (Fig. S18a). This is highly similar situation to LH2, except that the states have distinct energies.

Based on the excitonic state structure, Eq. 3 in the main text and also here above expects SE/GSB ratios of 0.21 for the weakly coupled chain, which is very close to the 1/5 limit, and 0.32 for the coupled chain, enhanced by the J-type coupling. In the F-2DES spectra (Fig. S18c), we find two diagonal peaks corresponding to the two chains, and two cross peaks indicating spectral correlations between them. The kinetics of the spectrally-integrated peaks is shown in Fig. S18b). The SE can be identified by the population transfer from the higher-energy chain (exponential decrease of the UDP amplitude in $t_2$) to the lower-energy chain (exponential increase of the LCP amplitude in $t_2$). Using equation (4) from the main text, we obtain from the UDP exponential decay the SE/GSB ratio of the higher-energy chain equal to 0.19, and using Eq. (5) we obtain from the LCP rise the SE/GSB ratio of the lower-energy chain equal to 0.30. These are in perfect agreement with the values expected from the state structure only, with the slightly lower values can be attributed to a marginal overlap of the peaks due to the long tails of the Lorentzian lineshapes. The prominence of the excited-state dynamics can thus be very well inferred from the state excitonic state structure, and identified in the F-2DES spectra as the dynamic component.

**F-2DES spectrum of LH2 at $t_2 = 0$ with AOPDF pulse shaper**

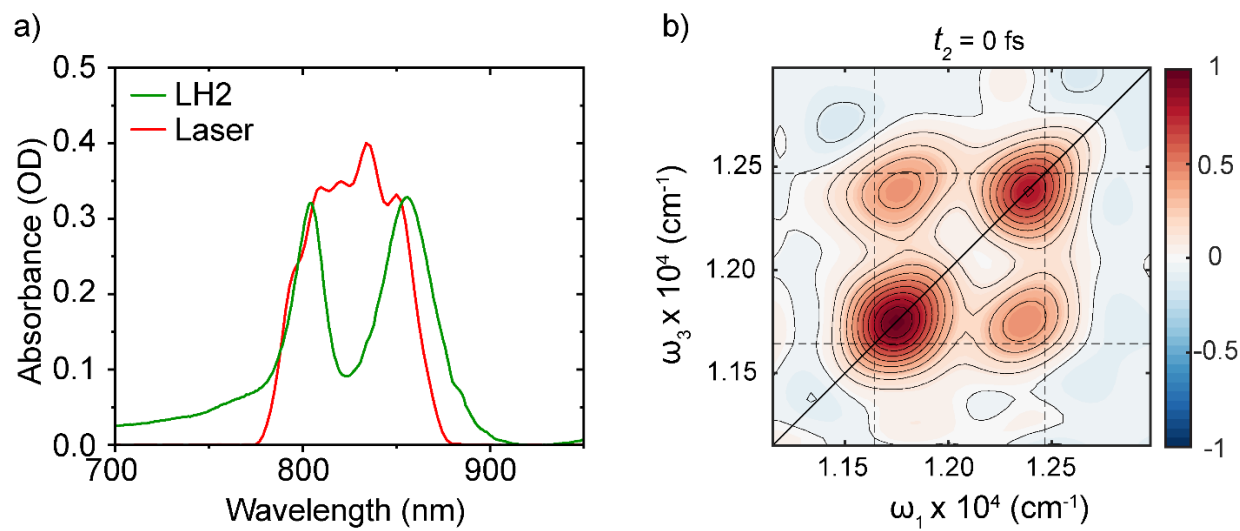

**Figure S19:** F-2DES experiment in a fully collinear geometry using the Dazzler. (a) Absorption spectra of LH2 along the used laser spectrum. (B) F-2D spectrum of LH2 at $t_2 = 0$.

We noticed that in our previous measurements of LH2 (10) as well as in other fluorescence-detected 2D spectra in literature of LH2 at $t_2 = 0$ (22), the cross peaks amplitude is asymmetric. In contrast to this, we report here symmetric cross peak amplitudes in the experiment as well as in the simulations. We identified two effects that might lead to asymmetric cross peak amplitudes. First, slightly different spectra for the four pulses might result in distorted peak amplitudes and second, imperfections during the phasing process. The latter can also cause a slight shift of the diagonal peaks away from the diagonal as observed in Fig. 1 of the main text. We investigated if indeed the described effects can be avoided. We measured the F-2DES spectrum of LH2 using an acousto-optic programmable dispersive filter (AOPDF, Dazzler) to create a four-pulse sequence and applied a 27-fold phase cycling scheme as reported in literature (23, 24). These experiments were performed at 1 kHz with the same light source as the C-2DES experiments, only using the pump beam path. For the F-2DES measurements with the Dazzler, the beam was focused to 50 µm at the sample position with 0.25 nJ pump energy. The emitted fluorescence was collected in the 90° geometry similar to the F-2DES at 1 MHz. Such a measurement ensures that phasing errors are small, and all laser pulses have identical spectra since all of them are created from the same beam using the pulse shaper. The laser spectrum along the absorption spectrum of LH2 is shown in Fig S19a. The resulting F-2DES spectrum at $t_2 = 0$ is shown in Fig. S19b. Since this laser spectrum has its maximum between the absorption maxima, the two main peaks are pulled along the diagonal to the center of the 2D spectrum, similar to how the LDP was pulled in the phase-modulated F-2DES spectra. We can clearly observe that the cross peaks have identical peak amplitudes, and the diagonal peaks are perfectly centered along the diagonal. Therefore, small differences in cross-peak amplitude at $t_2 = 0$ fs and slight shifts of the peaks off the diagonal can be attributed to small phasing errors and slight spectral differences since those effects are negligible in the implementation of F-2DES in which the Dazzler creates all four pulses. Note that none of these distortions of the F-2DES spectra will have any influence on the observed dynamics that we discuss in this paper.